\documentclass[%
 reprint,
 superscriptaddress,
 amsmath,amssymb,
 aps,
prl,
]{revtex4-2}

\usepackage{graphicx}% Include figure files
\usepackage{dcolumn}% Align table columns on decimal point
\usepackage{bm}% bold math
\usepackage{colortbl}
\usepackage[table]{xcolor}
\usepackage{makecell}
\usepackage[mathscr]{euscript}
\usepackage[table]{xcolor}
\usepackage{multirow}
\usepackage{braket}
\usepackage[pagewise]{lineno} 
\usepackage{pifont}

\begin{document}

%\preprint{APS/123-QED}

\title{Theory of polarization-switchable electrical conductivity anisotropy \\ in piezoelectric and ferroelectric semiconductors}

\author{Hong Jian Zhao}
 \affiliation{Key Laboratory of Material Simulation Methods and Software of Ministry of Education, College of Physics, Jilin University, Changchun 130012, China}
 \affiliation{Key Laboratory of Physics and Technology for Advanced Batteries (Ministry of Education), College of Physics, Jilin University, Changchun 130012, China}
 \affiliation{International Center of Future Science, Jilin University, Changchun 130012, China}
\author{Yanchao Wang}
 \email{wyc@calypso.cn}
 \affiliation{Key Laboratory of Material Simulation Methods and Software of Ministry of Education, College of Physics, Jilin University, Changchun 130012, China}
 \affiliation{State Key Laboratory of Superhard Materials, College of Physics, Jilin University, Changchun 130012, China}
\author{Laurent Bellaiche}
\affiliation{Smart Functional Materials Center, Physics Department and Institute for Nanoscience and Engineering, University of Arkansas, Fayetteville, Arkansas 72701, USA}
\affiliation{Department of Materials Science and Engineering, Tel Aviv University, Ramat Aviv, Tel Aviv 6997801, Israel}
\author{Yanming Ma}
\email{mym@jlu.edu.cn}
 \affiliation{Key Laboratory of Material Simulation Methods and Software of Ministry of Education, College of Physics, Jilin University, Changchun 130012, China}
 \affiliation{International Center of Future Science, Jilin University, Changchun 130012, China}
 \affiliation{State Key Laboratory of Superhard Materials, College of Physics, Jilin University, Changchun 130012, China}

\date{\today}

\begin{abstract}
The anisotropic propagation of particles is a fundamental transport phenomenon in solid state physics. As for crystalline semiconductors, the anisotropic charge transport opens novel designing routes for electronic devices, where the electrical or magnetic manipulation of anisotropic resistance provides essential guarantees. Motivated by the concept of anisotropic magnetoresistance, we develop an original theory on the electrically manipulatable anisotropic electroresistance. We show that piezoelectrics and ferroelectrics may showcase polarization-dependent anisotropic electrical conductivities between two perpendicular directions and the electrical conductivity anisotropy (ECA) is switchable by flipping the polarization. By symmetry analysis, we identify several point groups hosting the polarization-switchable ECA. These point groups simultaneously enable polarization-reversal induced conductivity change along specific directions, akin to the tunnelling electroresistance in ferroelectric tunnel junctions. First-principles-based conductivity calculations predict that piezoelectric AlP and ferroelectric KH$_2$PO$_4$ are two good semiconductors having such exotic charge transport. Our theory can motivate the design of intriguing anisotropic electronic devices (e.g., anisotropic memristor and field effect transistor).
\end{abstract}

\maketitle

\noindent
\section{I. Introduction}

The anisotropic transport in crystalline materials --- for instance, the anisotropic propagation of magnons~\cite{anisomagnon,anisomagnon2}, polaritons~\cite{anisopolar}, electrons or holes~\cite{eca2,amr1,afmspintronic2,anisospin,amr2,amr3,amr5,eca1,eca3,anisospin2} --- not only is of fundamental interest to solid state physics, but offers extensive development strategies for electronic, optoelectronic and spintronic devices~\cite{aniso,eca3,afmspintronic,amr4,afmspintronic2,anisospin2}. Mostly, the design of such devices relies on the electrical or magnetic manipulation of anisotropic transport~\cite{anisomagnon2,anisopolar,eca2,amr1,afmspintronic2,anisospin2}. A profound example is the anisotropic magnetoresistance that occurs in magnets, where longitudinal electrical resistance therein depends on the direction of magnetic order~\cite{afmspintronic,amr4,afmspintronic2,galvanomagnetic}. In ferromagnetic materials, magnetic field manipulates the magnetization direction, and changes the electrical resistance accordingly~\cite{afmspintronic,afmspintronic2}. This gives birth to early-stage magnetic recording devices~\cite{afmspintronic,amr4}, and yields magnetic sensors extensively deployed in automotive, aerospace, and biomedical realms~\cite{amr4}.

The anisotropic magnetoresistance is reminiscent of the conceptual anisotropic electroresistance. Previous works reveal the occurrence of tunnelling electroresistance, a counterpart of tunnelling magnetoresistance, in ferroelectric tunnel junctions, with the electroresistance characterized by the electrical resistance change induced by the reversal of electric polarization~\cite{ftj1,ftj2,ftj3,ftj4,ftj5,ftj6,ftj7,ftj8,ftj9}. With regard to anisotropic electroresistance, a particularly interesting research avenue is to discover crystalline semiconductors with polarization-switchable electrical conductivity anisotropy (ECA). Explicitly, it is desired to find semiconductors exhibiting $P$-dependent anisotropic conductivities between two orthogonal $\boldsymbol{\chi}$ and $\boldsymbol{\chi^\prime}$ directions [i.e., $\sigma_{\chi,\chi}(P) \neq \sigma_{\chi^\prime,\chi^\prime}(P)$], so that 
\begin{eqnarray}\label{eq:ecapol}
  \sigma_{\chi,\chi}(P) - \sigma_{\chi^\prime,\chi^\prime}(P)  = - \sigma_{\chi,\chi}(-P) + \sigma_{\chi^\prime,\chi^\prime}(-P),
\end{eqnarray}
where $P$ is an electric polarization induced by an electric field or spontaneously occurred.
Equation~(\ref{eq:ecapol}) guarantees that the ECA is manipulable and switchable by external electric field, which promises the design of anisotropic electronic devices (e.g., anisotropic memristor and field effect transistor)~\cite{aniso,eca2,eca1}. Besides that, Eq.~(\ref{eq:ecapol}) hints a possibility for a polarization-reversal induced conductivity change along $\boldsymbol{\chi}$ (or $\boldsymbol{\chi^\prime}$) direction in crystalline materials~\footnote{According to Eq.~(\ref{eq:ecapol}), the $\sigma_{\chi,\chi}(P) - \sigma_{\chi,\chi}(-P)$ conductivity difference along $\boldsymbol{\chi}$ direction equals $\sigma_{\chi^\prime,\chi^\prime}(P) + \sigma_{\chi^\prime,\chi^\prime}(-P) - 2\sigma_{\chi,\chi}(-P)$, which is not necessarily zero. The similar logic applies to the conductivity difference along $\boldsymbol{\chi^\prime}$ direction. In next section, we shall discuss this point in details.}, mimicking the tunnelling electroresistance in ferroelectric tunnel junctions. That is, semiconductors with polarization-switchable ECA may serve as alternates for ferroelectric tunnel junctions in device design.

The present work aims to provide guidelines for discovering the aforementioned elusive semiconductors. For this, we develop a polarization-dependent Boltzmann transport theory within effective mass approximation. Then, we perform symmetry analysis for crystallographic point groups, and extract the members that host polarization-switchable ECA. These point groups are associated with piezoelectric and ferroelectric materials. Simultaneously, our extracted point groups enable polarization-reversal driven conductivity change. 
This is further corroborated by first-principles-based electrical conductivity calculations, which helps to identify piezoelectric AlP and ferroelectric KH$_2$PO$_4$ as two representative semiconductors with such exotic transport phenomena. Our theory emphasizes the potential of various piezoelectric and ferroelectric semiconductors (including the celebrated zinc-blende type semiconductors) in designing devices based on anisotropic charge transport.  \\

\section{II. Methods}

We perform symmetry analyses, sketch drawings, and data plots with the help of Bilbao Crystallographic Server~\cite{aroyo2006,aroyo20062,aroyo2011}, SeeK-path~\cite{seekpath,seekpath2}, Mathematica~\cite{mma}, Matplotlib~\cite{hunter2007matplotlib}, Pymatgen~\cite{pymatgen}, and VESTA~\cite{momma2011vesta}. For numerical simulations, we employ the VASP~\cite{kresse1996efficient,kresse1999ultrasoft} code (based on PAW potentials~\cite{blochl1994projector}) to perform first-principles calculations and the Boltztrap2~\cite{boltztrap2} code to carry out charge transport calculations. 
We use the PBEsol exchange-correlation functional~\cite{pbesol} without involving spin-orbit interactions, and solve the following electronic configurations: $1s$ for H, $(2p,3s)$ for Na, $(3s,3p,4s)$ for K, $(4s,4p,5s)$ for Sr, $(3s,3p)$ for Al, $(2s,2p)$ for O and  $(3s,3p)$ for P. We select the kinetic energy cutoff values of 450 eV, 550 eV and 450 eV for AlP, KH$_2$PO$_4$, and NaSrP, respectively. The crystal structures of $F\bar{4}3m$ AlP, $Fdd2$ KH$_2$PO$_4$ and $P\bar{6}2m$ NaSrP are schematized in Fig.~1 and Fig.S2 of the Supplementary Material (SM)~\footnote{See Supplementary Material at {\color{blue}a link} which contains general remarks on polarization-switchable ECA, symmetry analysis on ECA, and some numerical results.}, where the SM also contains Refs.~\cite{thetafunction,nasrp2007}. In particular, we work with the super cell of AlP whose cell volume is twice that of its unit cell [see Fig.~1(a)]. The $k$-point meshes are $8\times8\times12$ for AlP, $6\times6\times10$ for KH$_2$PO$_4$, and $8\times8\times14$ for NaSrP for sampling the Brillouin zone, during the structural relaxations or band structure calculations. We carry out structural relaxations with the force convergence criterion of 1 meV/\AA, where the approach developed by Bellaiche and Fu~\cite{efield} enables the calculations of crystal structures under finite electric fields~\cite{chen2019electric,chen2016giant}. Starting from the relaxed crystal structures of AlP, KH$_2$PO$_4$ and NaSrP, we do self consistent field calculations to prepare the input files for the transport calculations; For accuracy, the $k$-point meshes are further increased to $60\times60\times90$ for AlP, $40\times40\times60$ for KH$_2$PO$_4$ and $60\times60\times104$ for NaSrP. The transport calculations are performed with a thermal smearing of 300 Kelvin, which yields non-zero conductivity values at the valence band maximum (VBM) or conduction band minimum (CBM). During the transport calculations, the crystal structures at various chemical potentials are kept fixed at their undoped structures. Even though free carriers create electrostatic screening, the polar distortions of a broad range of typical ferroelectrics well survive free carriers when the carrier concentration does not exceed $10^{20}$ cm$^{-3}$~\cite{doping} --- an upper limit associated with chemical doping~\cite{dopconcen}. This implies that a free carrier concentration of up to $10^{20}$ cm$^{-3}$ is insufficient to prevent the intrinsic or induced polar distortions in materials.

\section{III. Results and discussion}

\subsection{A. Theory of polarization-switchable ECA}

We start by developing the polarization-dependent Boltzmann transport theory. Under the approximation of constant relaxation time $\tau$, the longitudinal electrical conductivity (at zero Kelvin), contributed by a polarization-dependent $\epsilon(\mathbf{k},P)$ energy dispersion, is given by~\cite{drude1,drude2,drude3,drude4}
\begin{eqnarray}\label{eq:ohmicpol}
\frac{\sigma_{\chi,\chi}(P,\mu)}{\tau} \propto  \iiint \left[ \frac{\partial \epsilon(\mathbf{k},P)}{\partial k_\chi} \right]^2     \delta(\epsilon(\mathbf{k},P) - \mu)  \,d^3 \mathbf{k},
\end{eqnarray}
where $\delta$, $\mu$, $P$, and $\mathbf{k}\equiv k_x \mathbf{x} + k_y \mathbf{y} + k_z \mathbf{z}$~\footnote{We employ the $xyz$ Cartesian coordinate system with $x$, $y$, and $z$ axes being perpendicular to each other. As a convention, we use $\mathbf{x}$, $\mathbf{y}$, and $\mathbf{z}$ to denote the unit vectors along $x$, $y$, and $z$ axes, respectively.} denote Dirac's $\delta$ function, chemical potential, electric polarization, and electronic momentum, respectively. Writing the electrical conductivity as $\sigma_{\chi,\chi}(P,\mu)$ emphasizes its dependence on electric polarization and chemical potential.

To demonstrate our basic ideas, we work with a simplified energy dispersion 
\begin{eqnarray}\label{eq:disppol}
\epsilon(\mathbf{k},P)=\alpha_P k_x^2 + \beta_P k_y^2 + \gamma_P k_z^2,
\end{eqnarray}
with $\alpha_P$, $\beta_P$, and $\gamma_P$ depending on electric polarization. For simplicity too, we explore the ECA between the $\mathbf{x}$ and $\mathbf{y}$ directions by assuming that $\alpha_P$, $\beta_P$, and $\gamma_P$ are all positive numbers. The evaluation of Eqs.~(\ref{eq:ohmicpol}) and~(\ref{eq:disppol}) yields
\begin{eqnarray}\label{eq:condpol}
\frac{\sigma_{x,x}(P,\mu)}{\tau} \propto \frac{\alpha_P \mu^{\frac{3}{2}}}{\sqrt{\alpha_P \beta_P \gamma_P}}, \frac{\sigma_{y,y}(P,\mu)}{\tau} \propto \frac{\beta_P \mu^{\frac{3}{2}}}{\sqrt{\alpha_P \beta_P \gamma_P}}.
\end{eqnarray}
Up to first order in $P$, we expand $\alpha_P$, $\beta_P$, and $\gamma_P$ as $\alpha_{P} = \alpha_0 + \alpha_1 P$, $\beta_{P} = \beta_0 + \beta_1 P$, and $\gamma_{P} = \gamma_0 + \gamma_1 P$. This implies that 
\begin{eqnarray}\label{eq:condpol2}
\frac{\sigma_{x,x}(P,\mu)-\sigma_{y,y}(P,\mu)}{\tau} \propto \frac{(\alpha_0-\beta_0)+(\alpha_1-\beta_1)P}{\sqrt{\alpha_P \beta_P \gamma_P}}\mu^{\frac{3}{2}}.
\end{eqnarray}
To achieve the polarization-switchable ECA between the $\mathbf{x}$ and $\mathbf{y}$ directions, it is required that $\alpha_0 = \beta_0$, $\alpha_1\neq\beta_1$~\footnote{Our theory is established under the first-order approximation with respect to electric polarization $P$. Beyond such an approximation, this statement is not exactly true. For instance, $\alpha_P = \alpha_0 + \alpha_1 P + \alpha_3 P^3$ and $\beta_P = \alpha_0 + \beta_1 P + \beta_3 P^3$ may yield non-zero $\sigma_{x,x}(P,\mu)-\sigma_{y,y}(P,\mu)$, when $\alpha_1=\beta_1$ and $\alpha_3\neq\beta_3$.}, and $\alpha_P \beta_P \gamma_P = \alpha_{-P} \beta_{-P} \gamma_{-P}$ [see Eq.~(\ref{eq:ecapol})]. Similarly, we can derive the conditions for the polarizaiton-switchable ECA between the $\mathbf{y}$ and $\mathbf{z}$ directions and between the $\mathbf{z}$ and $\mathbf{x}$ directions. Within effective mass approximation, we provide more general discussion on polarization-switchable ECA in \textit{Section A} of the SM.

Excitingly, the polarization-switchable ECA implies the polarization-reversal induced conductivity change. For instance, the $\sigma_{x,x}(P,\mu)$ and $\sigma_{x,x}(-P,\mu)$ conductivity difference is given by
\begin{eqnarray}\label{eq:condpol3}
\frac{\sigma_{x,x}(P,\mu)-\sigma_{x,x}(-P,\mu)}{\tau} \propto \frac{2\alpha_1 P}{\sqrt{\alpha_P \beta_P \gamma_P}}\mu^{\frac{3}{2}},
\end{eqnarray}
which suggests the asymmetry for resistances between $+P$ and $-P$ states, with this asymmetry depending on the magnitude of electric polarization.\\

\subsection{B. Symmetry analysis}

We now provide guidelines to screen nonmagnetic~\footnote{As shown in Eqs.~(\ref{eq:disppol})--(\ref{eq:condpol3}), neither magnetism nor spin-orbit interaction is essential for polarization-switchable ECA. Magnetism and spin-orbit interaction are thus neglected in our derivations.} semiconductors enabling polarization-switchable ECA. 
To this end, we perform symmetry analysis for crystallographic point groups. We first work with nonpolar point groups and extend our discussion to polar point groups at a later stage. Depending on polarization $\mathbf{P}\equiv P_x \mathbf{x} + P_y \mathbf{y} + P_z \mathbf{z}$, the formalism of the energy dispersion is written as
\begin{eqnarray}\label{eq:heffmasspol}
\epsilon(\mathbf{k},\mathbf{P})  = \sum_{u,v=x,y,z}\tau_{uv}k_u k_v + \sum_{u,v,w=x,y,z}\lambda_{wuv} P_w k_u k_v,
\end{eqnarray}
where $\tau_{uv}$ and $\lambda_{wuv}$ coefficients characterize the bare effective masses and polarization-dressed effective masses of the carriers, respectively. Within effective mass approximation, Eq.~(\ref{eq:heffmasspol}) is derived around the Brillouin Zone center and up to first order in $P_x$, $P_y$ and $P_z$.

\begin{table}[ht]
\renewcommand{\arraystretch}{1.5}
\caption{\label{tab:ptbar4m2pol} The $\Gamma_1$ and $\Gamma_4$ irreducible representations for the $\bar{4}m2$ point group~\cite{point,koster,space}. In this table, $\mathfrak{1}$, $\bar{\mathfrak{4}}_z^{+}$, and $\bar{\mathfrak{4}}_z^{-}$ are identity operation, four-fold rotoinversion along $\mathbf{z}$ direction, and the inverse of $\bar{\mathfrak{4}}_z^{+}$, respectively; $\mathfrak{m}_x$ ($\mathfrak{m}_y$) is the mirror plane perpendicular to $\mathbf{x}$ ($\mathbf{y}$) direction, and $\mathfrak{2}_{xy}$ ($\mathfrak{2}_{x\bar{y}}$) is the two-fold rotation along $\mathbf{x}+\mathbf{y}$ ($\mathbf{x}-\mathbf{y}$) direction.} 
\begin{ruledtabular}
\begin{tabular}{ccccccc}
 & $\mathfrak{1}$ & $\mathfrak{2}_z$  & $\bar{\mathfrak{4}}_z^{+}$, $\bar{\mathfrak{4}}_z^{-}$ &  $\mathfrak{2}_{xy}$, $\mathfrak{2}_{x\bar{y}}$ &  $\mathfrak{m}_{x}$, $\mathfrak{m}_{y}$  &  Bases \\
 \hline 
$\Gamma_1$ & $+1$ & $+1$ & $+1$ & $+1$ & $+1$ &  $k_x^2 + k_y^2$, $k_z^2$     \\
$\Gamma_4$ & $+1$ & $+1$ & $-1$ & $-1$ & $+1$ &  $P_z$, $k_x^2 - k_y^2$     \\
\end{tabular}
\end{ruledtabular}
\end{table}

In the following, we use $\bar{4}m2$ and $\bar{4}2m$ point groups to exemplify the derivations of the $P_z$-dependent energy dispersions. Our derivations are based on the group representation theory (see e.g., Refs.~\cite{gtpack,symmetry}).
As shown in Table~\ref{tab:ptbar4m2pol}, the $k_x^2+k_y^2$ and $k_z^2$ bases belong to the identity representation $\Gamma_1$. This indicates that $(k_x^2+k_y^2)$ and $k_z^2$ are invariant with respect to the $\bar{4}m2$ point group, yielding $\tau_{xx}(k_x^2+k_y^2)$ and $\tau_{zz}k_z^2$ terms. The bases for the $\Gamma_4$ representation are $P_z$ and $k_x^2-k_y^2$, which are not invariant (e.g., $\bar{\mathfrak{4}}_z^{+}: P_z\rightarrow -P_z, k_x^2-k_y^2 \rightarrow -k_x^2+k_y^2$). Since $\Gamma_4\otimes\Gamma_4=\Gamma_1$, we can combine $P_z$ with $k_x^2-k_y^2$ to generate a new base for the $\Gamma_1$ representation [i.e., $\lambda_{zxx}P_z(k_x^2-k_y^2)$]. Overall, the $P_z$-dependent energy dispersion associated with $\bar{4}m2$ is $\epsilon(\mathbf{k},P_z)  = (\tau_{xx}+\lambda_{zxx}P_z) k_x^2 + (\tau_{xx}-\lambda_{zxx}P_z) k_y^2 + \tau_{zz}k_z^2$. According to Eq.~(\ref{eq:condpol2}), this suggests the polarization-switchable ECA between the $\mathbf{x}$ and $\mathbf{y}$ directions. Similarly, we can derive the $P_z$-dependent energy dispersion for the $\bar{4}2m$ point group as $\epsilon(\mathbf{k},P_z)  =  \tau_{xx}(k_x^2+k_y^2)+\tau_{zz}k_z^2 + \lambda_{zxy}P_z k_x k_y$ [see ``\textit{The $\bar{4}2m$ point group}'' in \textit{Section B} of our SM]. Such an energy dispersion implies the polarization-switchable ECA between the $\mathbf{x}+\mathbf{y}$ and $-\mathbf{x}+\mathbf{y}$ directions. To show this, we define an $x^\prime y^\prime z$ Cartesian coordinate system whose basis vectors are $\mathbf{x}^\prime=(\mathbf{x}+\mathbf{y})/\sqrt{2}$, $\mathbf{y}^\prime=(-\mathbf{x}+\mathbf{y})/\sqrt{2}$, and $\mathbf{z}$ (see Fig.~S1 of the SM). In $x^\prime y^\prime z$ coordinate system, the electronic momentum is written as $\mathbf{k}=k_{x^\prime} \mathbf{x}^\prime + k_{y^\prime} \mathbf{y}^\prime + k_z \mathbf{z}$, and the energy dispersion associated with $\bar{4}2m$ becomes $\epsilon(\mathbf{k},P_z)  =  (\tau_{xx} + \lambda_{zxy} P_z/2) k_{x^\prime}^2 +  (\tau_{xx} - \lambda_{zxy} P_z/2) k_{y^\prime}^2 + \tau_{zz}k_z^2$. Therefore, the $\bar{4}2m$ point group hosts a polarization-switchable ECA between $\mathbf{x}^\prime$ and $\mathbf{y}^\prime$ (i.e., $\mathbf{x}+\mathbf{y}$ and $-\mathbf{x}+\mathbf{y}$).

\begin{table}[ht]
\caption{\label{tab:pointeca} Point groups enabling the polarization-switchable ECA, derived up to first order in electric polarization $P_w$ ($w=x,y,z$). The entries below each $P_w$ are endowed with $(\chi,\chi^\prime)$ if ECA between the $\boldsymbol{\chi}$ and $\boldsymbol{\chi^\prime}$ directions is enabled by gaining $P_w$ and is switchable by reversing $P_w$. Otherwise, the entries are endowed with ``$\cdots$''. Here, $\chi$ or $\chi^\prime$ is written as $u$, $uv$, or $\bar{u}v$ ($u,v$ being $x$, $y$, or $z$), where $uv$ and $\bar{u}v$ denote the $\mathbf{u}+\mathbf{v}$ and $-\mathbf{u}+\mathbf{v}$ directions, respectively.} 
\begin{ruledtabular}
\begin{tabular}{lccc}
      \multicolumn{4}{c}{{\textit{Category A: rigorous cases}}}  \\
 & $P_x$ & $P_y$ & $P_z$  \\ \cline{2-4}
$\bar{4}$ & $\cdots$ & $\cdots$ & $(x,y)$; $(xy,\bar{x}y)$ \\
$\bar{4}2m$ & $\cdots$ & $\cdots$ & $(xy,\bar{x}y)$ \\
$\bar{4}m2$ & $\cdots$ & $\cdots$ & $(x,y)$ \\
$312$  &  $(xy,\bar{x}y)$ & $\cdots$ & $\cdots$ \\
$321$  &  $\cdots$  &   $(xy,\bar{x}y)$  & $\cdots$ \\
$\bar{6}m2$ &  $(xy,\bar{x}y)$ & $\cdots$  & $\cdots$ \\
$\bar{6}2m$ & $\cdots$ &  $(xy,\bar{x}y)$ & $\cdots$ \\
$23$ & $(yz,\bar{y}z)$   &   $(zx,\bar{z}x)$  &   $(xy,\bar{x}y)$  \\
$\bar{4}3m$ & $(yz,\bar{y}z)$   &   $(zx,\bar{z}x)$  &   $(xy,\bar{x}y)$  \\ 
 \hline\hline   
\multicolumn{4}{c}{{\textit{Category B: approximate cases}}}   \\
 & $P_x$ & $P_y$ & $P_z$  \\  \cline{2-4}
$\bar{6}$ & $(x,y)$; $(xy,\bar{x}y)$ & $(x,y)$; $(xy,\bar{x}y)$ &  $\cdots$  \\
$\bar{6}m2$ & $\cdots$  & $(x,y)$  & $\cdots$ \\
$\bar{6}2m$ & $(x,y)$ &  $\cdots$ & $\cdots$ \\
$23$ & $(\bar{y}z,x)$; $(x,yz)$   &   $(\bar{z}x,y)$; $(y,zx)$  &   $(\bar{x}y,z)$; $(z,xy)$  \\
$\bar{4}3m$ & $(\bar{y}z,x)$; $(x,yz)$   &   $(\bar{z}x,y)$; $(y,zx)$  &   $(\bar{x}y,z)$; $(z,xy)$  \\
\end{tabular}
\end{ruledtabular}
\end{table}

Our analyses for the entire nonpolar crystallographic point groups are shown in \textit{Section B} and \textit{Section C} of our SM. This helps to identify a variety of point groups that enable the polarization-switchable ECA, as summarized in Table~\ref{tab:pointeca}. These point groups are associated with nonpolar piezoelectric materials. The cases listed in Table~\ref{tab:pointeca} can be classified into \textit{Category A} and \textit{Category B}. Belonging to \textit{Category A} are cases where for ECA between the $\boldsymbol{\chi}$ and $\boldsymbol{\chi}^\prime$ directions (driven by $P_w$ polarization) the point group symmetry protects the $\sigma_{\chi,\chi}(P_w,\mu)-\sigma_{\chi^\prime,\chi^\prime}(P_w,\mu)=-\sigma_{\chi,\chi}(-P_w,\mu)+\sigma_{\chi^\prime,\chi^\prime}(-P_w,\mu)$ relationship. In such sense, the polarization-switchable feature for the ECA is rigorous, being robust against higher-order corrections~\footnote{By higher-order corrections, we mean corrections that go beyond effective mass approximation or involve higher-order contributions from electric polarization. Such higher-order corrections should be compatible with the corresponding point group symmetry.}.  
For instance, the $\bar{4}3m$ point group contains two-fold rotation symmetry operations along the $\mathbf{x}$ and $\mathbf{y}$ directions. These symmetry operations protect the polarization-switchable feature for the $P_z$-induced ECA between the $\mathbf{x}+\mathbf{y}$ and $-\mathbf{x}+\mathbf{y}$ directions, because they provide a linkage between $\sigma_{xy,xy}(P_z,\mu)-\sigma_{\bar{x}y,\bar{x}y}(P_z,\mu)$ and $\sigma_{\bar{x}y,\bar{x}y}(-P_z,\mu)-\sigma_{xy,xy}(-P_z,\mu)$. The \textit{Category B} contains cases for which the polarization-switchable ECA, derived within effective mass approximation (up to first-order in polarization), is not rigorously guaranteed by point group symmetry. As an example, the $\bar{6}2m$ point group hosts an ECA between the $\mathbf{x}$ and $\mathbf{y}$ directions driven by $P_x$ polarization, while this point group has no symmetry operations linking $P_x$ with $-P_x$ (see \textit{Section C} of the SM). This means that the $\sigma_{x,x}(P_x,\mu)-\sigma_{y,y}(P_x,\mu)=\sigma_{y,y}(-P_x,\mu)-\sigma_{x,x}(-P_x,\mu)$ relationship associated with the $\bar{6}2m$ point group may be broken when involving higher-order corrections such as $\zeta_{xyz}^{lmn} P_x^l k_x^m k_y^n$ ($l,m,n$ being natural numbers with $l\ge2$ or $m+n>2$).

Our aforementioned polarization-switchable ECAs in nonpolar piezoelectrics are volatile, that is, an electric field is required to maintain the ECAs. A ferroelectric has a spontaneous electric polarization, and the reversal of the polarization involves an intermediate phase. If the point group of this intermediate phase belongs to the cases listed in Table~\ref{tab:pointeca}, the ferroelectric may host ferroelectrically switchable ECA (i.e., nonvolatile). For instance, a nonpolar phase with $\bar{4}m2$ point group showcases $P_z$-switchable ECA between the $\mathbf{x}$ and $\mathbf{y}$ directions (see Table~\ref{tab:pointeca}). By gaining a spontaneous $P_z$ polarization, this phase becomes ferroelectric with the $mm2$ point group, and such a ferroelectric phase enables ferroelectrically switchable $\sigma_{x,x}(P_z,\mu) - \sigma_{y,y}(P_z,\mu)$ anisotropic conductivity. Following this logic, we derive guidelines for screening ferroelectrics with ferroelectrically switchable ECA (summarized in Table~\ref{tab:ferroeca}).

\begin{table}[ht]
\renewcommand{\arraystretch}{1.5}
\caption{\label{tab:ferroeca} Ferroelectrically switchable ECA. The first column lists the point groups for  intermediate phases. Gaining spontaneous polarizations drive these phases ferroelectric, and the resulting polar point groups are listed in the second column; Each spontaneous polarization is shown as $P_x$, $P_y$, or $P_z$ in the parenthesis. The $(\chi,\chi^\prime)$ in the third column indicates the ferroelectrically switchable ECA between $\boldsymbol{\chi}$ and $\boldsymbol{\chi^\prime}$. The subscripts (if existing) in the polar point group symbols indicate the directions of the associated symmetry operations. To be specific, $\alpha$ in $2_\alpha$ labels the direction for the twofold rotation axis, and $\beta$ in $m_\beta$ marks the normal direction for the mirror plane ($\alpha$ and $\beta$ being $x$, $y$, $z$, $\cdots$).} 
\begin{ruledtabular}
\begin{tabular}{lcc}
\multicolumn{3}{c}{\textit{Category A: rigorous cases}} \\
Intermediate       &    Ferroelectric       &   ECA      \\
\hline
$\bar{4}$       &   $2_z$ ($P_z$)       &  $(x,y)$; $(xy,\bar{x}y)$      \\
$\bar{4}2m$     &   $m_{xy} m_{x\bar{y}} 2_z$  ($P_z$)   &    $(xy,\bar{x}y)$\\
$\bar{4}m2$     &    $m_{x} m_{y} 2_z$  ($P_z$)   &   $(x,y)$  \\
$312$           &         $1$        ($P_x$)       &   $(xy,\bar{x}y)$  \\
$321$           &         $1$        ($P_y$)       &   $(xy,\bar{x}y)$   \\
$\bar{6}m2$     &       $m_z$  ($P_x$)               &   $(xy,\bar{x}y)$ \\
$\bar{6}2m$     &       $m_z$  ($P_y$)               &   $(xy,\bar{x}y)$  \\
$23$            &       $2_x$  ($P_x$)               &   $(yz,\bar{y}z)$   \\
$23$            &       $2_y$  ($P_y$)               &   $(zx,\bar{z}x)$   \\
$23$            &       $2_z$  ($P_z$)               &   $(xy,\bar{x}y)$   \\
$\bar{4}3m$     &       $m_{yz} m_{y\bar{z}}2_x$  ($P_x$)               &   $(yz,\bar{y}z)$   \\
$\bar{4}3m$     &       $m_{zx} m_{z\bar{x}}2_y$  ($P_y$)               &   $(zx,\bar{z}x)$   \\
$\bar{4}3m$     &       $m_{xy} m_{x\bar{y}}2_z$  ($P_z$)               &   $(xy,\bar{x}y)$   \\
\hline
\hline
\multicolumn{3}{c}{\textit{Category B: approximate cases}} \\
Intermediate       &    Ferroelectric       &   ECA      \\
\hline
$\bar{6}$      &        $m_z$   ($P_x$)        &    $(x,y)$; $(xy,\bar{x}y)$   \\       
$\bar{6}$      &        $m_z$   ($P_y$)        &    $(x,y)$; $(xy,\bar{x}y)$   \\   
$\bar{6}m2$     &       $m_z m_x 2_y$  ($P_y$)               &  $(x,y)$ \\
$\bar{6}2m$     &       $m_y m_z 2_x$  ($P_x$)               &  $(x,y)$ \\
$23$            &       $2_x$  ($P_x$)               &   $(\bar{y}z,x)$; $(x,yz)$   \\
$23$            &       $2_y$  ($P_y$)               &   $(\bar{z}x,y)$; $(y,zx)$   \\
$23$            &       $2_z$  ($P_z$)               &   $(\bar{x}y,z)$; $(z,xy)$   \\
$\bar{4}3m$     &        $m_{yz} m_{y\bar{z}}2_x$  ($P_x$)      &   $(\bar{y}z,x)$; $(x,yz)$   \\
$\bar{4}3m$     &        $m_{zx} m_{z\bar{x}}2_y$ ($P_y$)        &   $(\bar{z}x,y)$; $(y,zx)$   \\
$\bar{4}3m$     &        $m_{xy} m_{x\bar{y}}2_z$ ($P_z$)        &   $(\bar{x}y,z)$; $(z,xy)$   \\
\end{tabular}
\end{ruledtabular}
\end{table}

To end this section, we comment on the correlation between polarization-switchable ECA and polarization-reversal driven conductivity change. In \textit{Section A} of the SM, we show that the $P_w$-switchable ECA between the $\boldsymbol{\chi}$ and $\boldsymbol{\chi^\prime}$ directions is rooted in $(\eta_{\chi\chi}+\kappa_{w\chi\chi}P_w)k_\chi^2$ or $(\eta_{\chi\chi}-\kappa_{w\chi\chi}P_w)k_{\chi^\prime}^2$ term. According to Eqs.~(\ref{eq:disppol})--(\ref{eq:condpol3}), $(\eta_{\chi\chi}+\kappa_{w\chi\chi}P_w)k_\chi^2$ and $(\eta_{\chi\chi}-\kappa_{w\chi\chi}P_w)k_{\chi^\prime}^2$ terms imply the polarization-reversal driven conductivity changes along the $\boldsymbol{\chi}$ and $\boldsymbol{\chi^\prime}$ directions, respectively. As for Table~\ref{tab:pointeca}, we conclude that every $(\chi,\chi^\prime)$ entry below $P_w$ indicates the conductivity changes resulting from the reversal of $P_w$, that is, $\sigma_{\chi,\chi}(P_w,\mu)-\sigma_{\chi,\chi}(-P_w,\mu)\neq0$ and $\sigma_{\chi^\prime,\chi^\prime}(P_w,\mu)-\sigma_{\chi^\prime,\chi^\prime}(-P_w,\mu)\neq0$. The only exceptions are associated with $23$ and $\bar{4}3m$ point groups, where $\sigma_{w,w}(P_w,\mu)-\sigma_{w,w}(-P_w,\mu)=0$ holds for $w$ being $x$, $y$, or $z$. These arguments are readily extended to the cases shown in Table~\ref{tab:ferroeca}. \\

\begin{figure}[t!]
\centering
\includegraphics[width=0.8\linewidth]{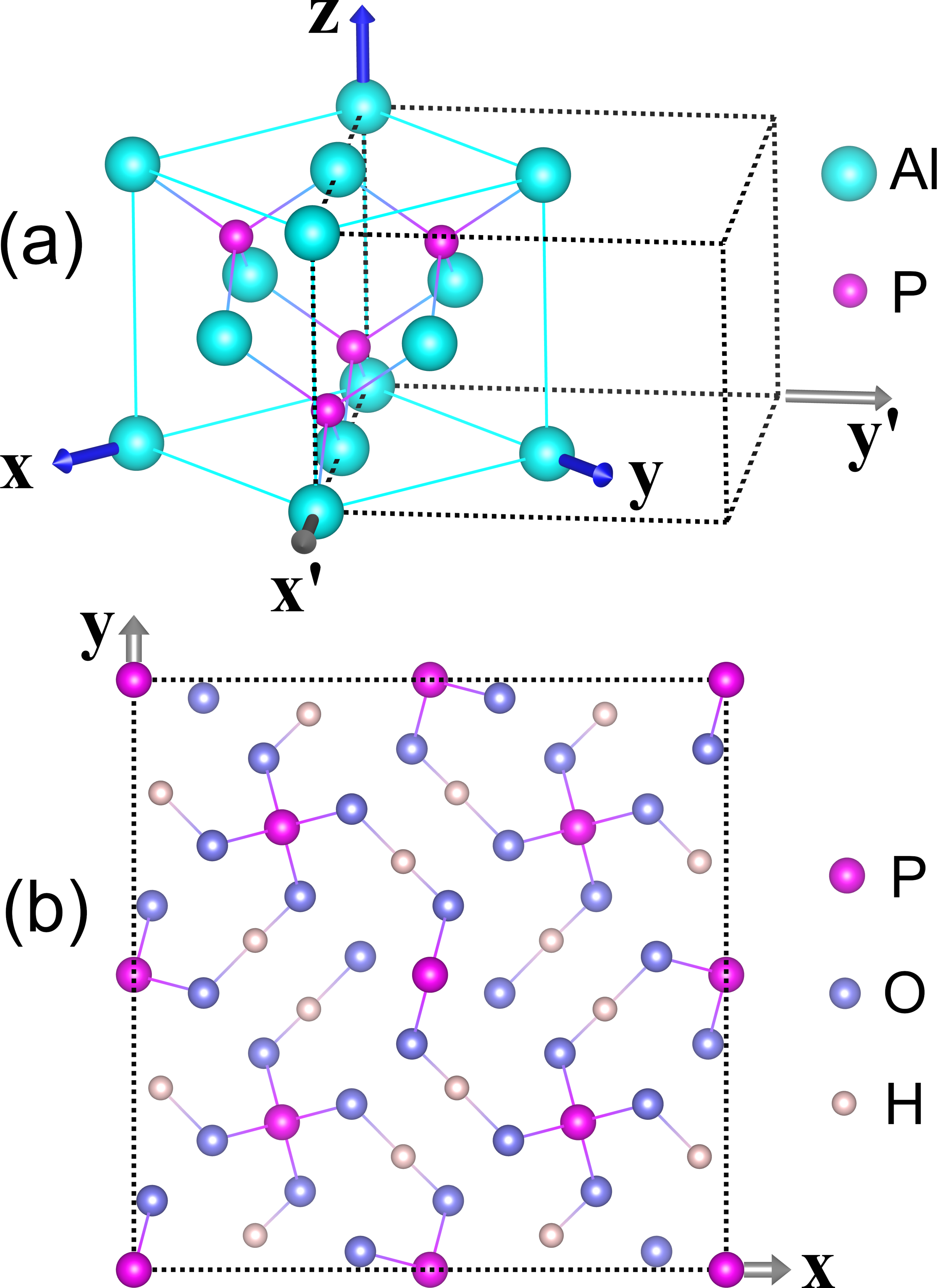}
\caption{\label{fig:crystsketch} Schematizations of the crystal structures for $\bar{4}3m$ AlP (a) and $mm2$ KH$_2$PO$_4$ (b). The $\mathbf{x}$, $\mathbf{y}$, and $\mathbf{z}$ unit vectors in both panels are perpendicular to each other [$\mathbf{z}$ in panel (b) being omitted]. In panel (a), $\mathbf{x^\prime}$ and $\mathbf{y^\prime}$ are defined as $\mathbf{x^\prime}=(\mathbf{x}+\mathbf{y})/\sqrt{2}$ and $\mathbf{y^\prime}=(-\mathbf{x}+\mathbf{y})/\sqrt{2}$. The boxes enclosed by cyan solid lines (i.e., $xyz$ coordinate system) and black dashed lines (i.e., $x^\prime y^\prime z$ coordinate system) denote the unit cell of AlP and the super cell used in our simulations, respectively. In panel (b), the K ions are not shown.}
\end{figure}

\subsection{C. Semiconductors with polarization-switchable ECA}
We move on to identify semiconductors with polarization-switchable ECA, by electrical conductivity calculations based on first principles. As shown in Tables~\ref{tab:pointeca} and~\ref{tab:ferroeca}, a variety of point groups host polarization-dependent ECAs that are rigorously switched by flipping the polarization~\footnote{Tables~\ref{tab:pointeca} and~\ref{tab:ferroeca} also contain cases associated with approximately polarization-switchable ECA. We demonstrate and verify this by computing the electrical conductivities for the NaSrP semiconductor, which is shown in~\textit{Section D} of our SM.}. We select $\bar{4}3m$ in Table~\ref{tab:pointeca} and $\bar{4}m2$ in Table~\ref{tab:ferroeca} as our platforms, where the former is the point group of various zinc-blende type semiconductors (e.g., AlP~\cite{alpcryst}) being technologically important~\cite{zinckresse,zinckresse2} and the latter is the point group of the intermediate paraelectric phase for ferroelectric KH$_2$PO$_4$~\cite{kdp1,kdp2,kdp3}.

\begin{figure}[t!]
\centering
\includegraphics[width=1\linewidth]{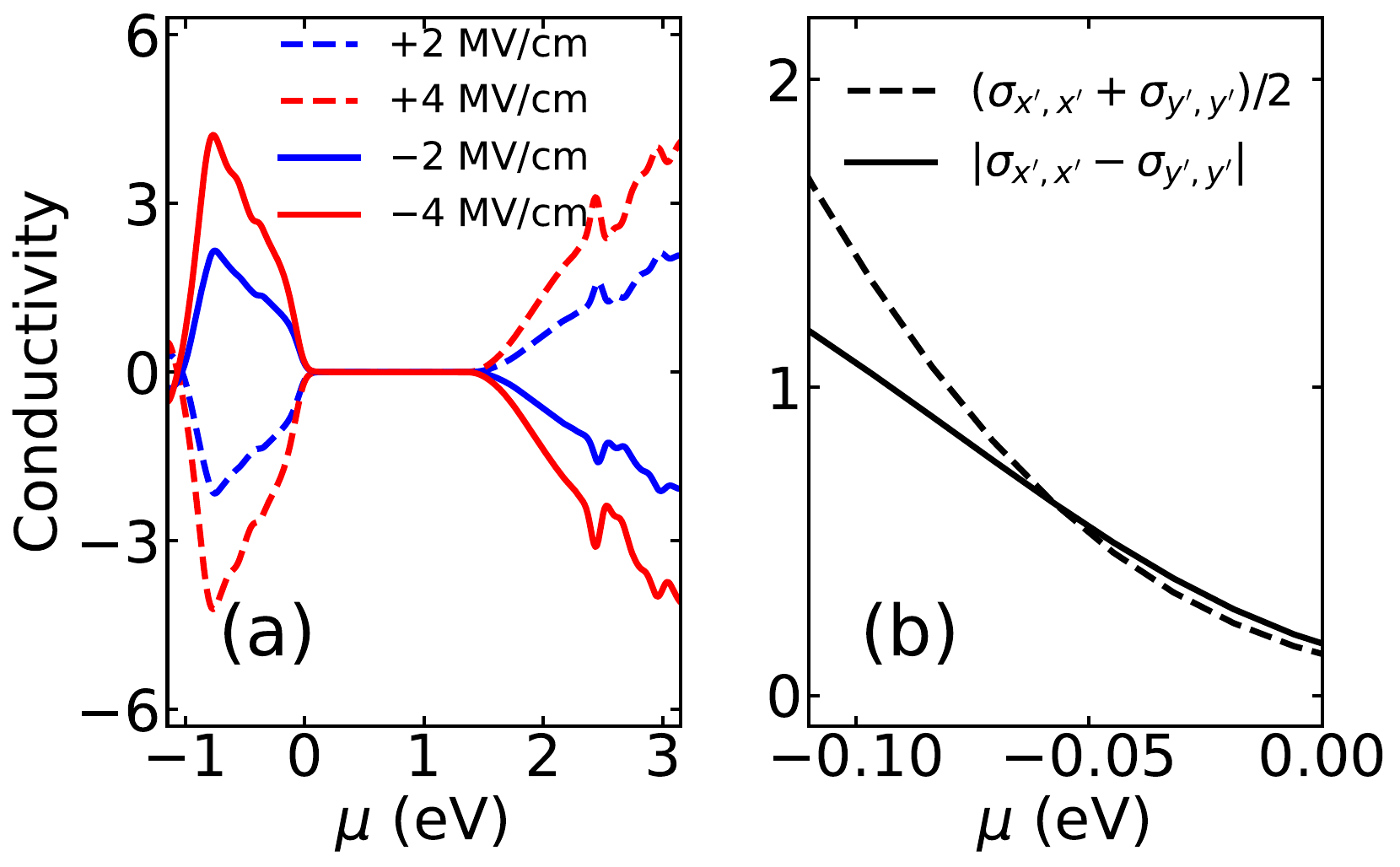}
\caption{\label{fig:condalp} {The ECA in AlP as functions of chemical potential $\mu$, where $\mu=0$ is referred to as the corresponding valence band maximum. Panel (a): $(\sigma_{x^\prime,x^\prime}-\sigma_{y^\prime,y^\prime})$ driven by electric fields along $\pm\mathbf{z}$ direction. Panel (b): $(\sigma_{x^\prime,x^\prime}+\sigma_{y^\prime,y^\prime})/2$ and $|\sigma_{x^\prime,x^\prime}-\sigma_{y^\prime,y^\prime}|$ associated with an electric field of $+4$ MV/cm along the $\mathbf{z}$ direction. The unit for the vertical axes is $10^{19}\tau/(\Omega\cdot m\cdot s)$.}}
\end{figure}

We use AlP and KH$_2$PO$_4$ to validate our theory. In Fig.~\ref{fig:crystsketch}(a), the box enclosed by cyan solid lines schematizes the unit cell of AlP (i.e., $xyz$ coordinate system), while the larger super cell (i.e., $x^\prime y^\prime z$ coordinate system) is the one employed in our numerical simulations. To polarize the AlP semiconductor, we apply external electric fields of $\pm2$ and $\pm4$ MV/cm along $\mathbf{z}$ direction to such a material. This creates four different $P_z$ polarization states in AlP, and yields finite $(\sigma_{x^\prime,x^\prime}-\sigma_{y^\prime,y^\prime})$ conductivity differences being $P_z$ dependent [see Fig.~\ref{fig:condalp}(a)]. According to Fig.~\ref{fig:crystsketch}(a), such conductivity differences characterize the ECA between the $\mathbf{x}+\mathbf{y}$ and $-\mathbf{x}+\mathbf{y}$ directions. The conductivity differences associated with electric fields of $+4$ and $-4$ MV/cm ($+2$ and $-2$ MV/cm) are rather symmetric with respect to the horizontal axis at $(\sigma_{x^\prime,x^\prime}-\sigma_{y^\prime,y^\prime})=0$. This confirms our theory that $P_z$ drives strictly polarization-switchable ECA in materials with $\bar{4}3m$ point group. Fig.~\ref{fig:condalp}(b) shows $\Delta \sigma=|\sigma_{x^\prime,x^\prime}-\sigma_{y^\prime,y^\prime}|$ and $\sigma_0=(\sigma_{x^\prime,x^\prime}+\sigma_{y^\prime,y^\prime})/2$ for AlP, associated with an electric field of $+4$ MV/cm along $\mathbf{z}$. Strikingly, $\Delta \sigma$ is sizable as compared to $\sigma_0$, in the vicinity of valence band maximum. We use $\Delta \sigma/\sigma_0$ to describe the relative anisotropic conductivity, and this resembles the approach for quantifying the relative anisotropic magnetoresistance (see e.g., Ref.~\cite{amr4}). The $\Delta \sigma/\sigma_0$ ranges from $\sim125\%$ to $\sim75\%$ when decreasing the chemical potential $\mu$ from $0.0$ to $-0.1$ eV. Experimentally, such a variation of $\mu$ is achievable by doping AlP with Be which creates $p$-type hole carriers~\cite{alpdopingbe}.

\begin{figure}[t!]
\centering
\includegraphics[width=1\linewidth]{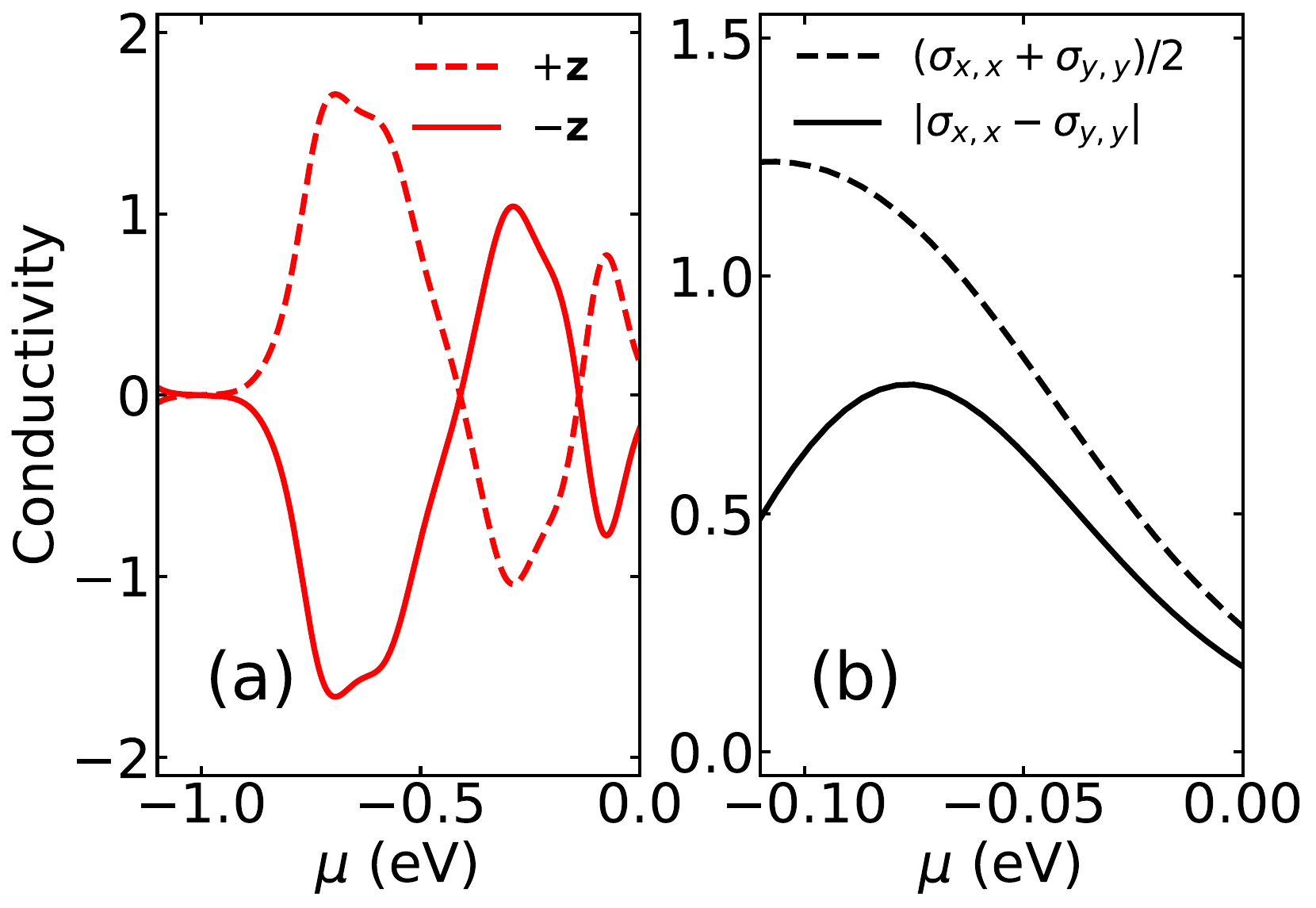}
\caption{\label{fig:condkdp} {The ECA in ferroelectric KH$_2$PO$_4$ as functions of chemical potential $\mu$, where $\mu=0$ is referred to as the corresponding valence band maximum. Panel (a): $(\sigma_{x,x}-\sigma_{y,y})$ associated with electric polarizations along $\pm \mathbf{z}$ direction. Panel (b): $(\sigma_{x,x}+\sigma_{y,y})/2$ and $|\sigma_{x,x}-\sigma_{y,y}|$ associated with an electric polarization along the $+\mathbf{z}$ direction. The unit for the vertical axes is $10^{19}\tau/(\Omega\cdot m\cdot s)$.}}
\end{figure}

Ferroelectric KH$_2$PO$_4$ has a point group of $mm2$~\cite{kdp1,kdp2,kdp3} and a spontaneous polarization along $\mathbf{z}$ [see Fig.~\ref{fig:crystsketch}(b)]. The point group of the intermediate paraelectric phase, for polarization reversal, is $\bar{4}m2$~\cite{kdp1,kdp2,kdp3}. According to Table~\ref{tab:ferroeca}, KH$_2$PO$_4$ exhibits ferroelectrically switchable anisotropic conductivity between $\mathbf{x}$ and $\mathbf{y}$, which stems from spontaneous polarization and hence are nonvolatile. Our assertions are validated by Fig.~\ref{fig:condkdp}(a). As shown in Fig.~\ref{fig:condkdp}(b), $\Delta \sigma=|\sigma_{x,x}-\sigma_{y,y}|$ is sizable compared with $\sigma_0=(\sigma_{x,x}+\sigma_{y,y})/2$. In the range of $-0.1~\mathrm{eV} \le \mu \le 0~\mathrm{eV}$, the relative anisotropic conductivity $\Delta \sigma/\sigma_0$ varies between $50\%$ and $76\%$.

We now check the conductivity changes in AlP and KH$_2$PO$_4$ that are induced by the reversal of polarization. We polarize AlP by applying $\pm4$ MV/cm electric fields along the $\mathbf{z}$ direction. As shown in Fig.~S5 of the SM, this yields $\pm P_z$ polarization states with unequal $\sigma_1=\sigma_{x^\prime,x^\prime}(P_z,\mu)$ and $\sigma_2=\sigma_{x^\prime,x^\prime}(-P_z,\mu)$ conductivities. Following the definition of tunneling magnetoresistance ratio~\cite{mtjmr2}, we define our conductivity change ratio as $|\sigma_1-\sigma_2|/\sigma_\mathrm{min}$, where $\sigma_\mathrm{min}$ is the minimum between $\sigma_1$ and $\sigma_2$. At $\mu=0~\mathrm{eV}$, the conductivity change ratio in AlP reaches $\sim 333\%$. As for KH$_2$PO$_4$, switching the direction of its polarization between $+\mathbf{z}$ and $-\mathbf{z}$ yields diverse $\sigma_1=\sigma_{x,x}(P_z,\mu)$ and $\sigma_2=\sigma_{x,x}(-P_z,\mu)$ conductivities (see Fig.~S6 of the SM), with a conductivity change ratio of $\sim104\%$ at $\mu=0~\mathrm{eV}$.\\

\section{IV. Summary and outlook}

Within effective mass approximation, we have developed a theory on polarization-switchable ECA in semiconductors, where the ECA exhibits a first-order response to electric polarization and is reversible by flipping the polarization. We have shown that the polarization-switchable ECA is accompanied with polarization-reversal induced conductivity change. This may occur in piezoelectrics or ferroelectrics with their possible crystallographic point groups summarized in Tables~\ref{tab:pointeca} and~\ref{tab:ferroeca}. Our theory is validated by electrical conductivity calculations based on first principles. This yields the predictions of piezoelectric AlP and ferroelectric KH$_2$PO$_4$ as two representative semiconductors with the aforementioned transport phenomena.

In the vicinity of the valence band maximum, our estimated relative anisotropic conductivities in AlP (driven by $\pm 4$ MV/cm electric fields) and KH$_2$PO$_4$ (occurred spontaneously) exceed $50\%$. Such values are not less than the relative anisotropic magnetoresistance in permalloy (e.g., between $16\%$ and $25\%$ for Ni$_{80}$Fe$_{20}$~\cite{amr4}) that is used in industry. At $\mu=0$ eV, AlP and KH$_2$PO$_4$ exhibit polarization-reversal induced conductivity changes with change ratios of $\sim333\%$ and $\sim104\%$, respectively. The ratio in KH$_2$PO$_4$ is lower than the magnetoresistance ratio in commercial magnetic tunnel junctions (usually, larger than $250\%$)~\cite{mtjmr}, while the ratio in AlP exceeds $250\%$. In view of this, polarization-switchable electrical conductivity anisotropy and polarization-reversal induced conductivity change in AlP should be promising for device design. As an outlook, we expect that our work opens a route toward next-generation electronic devices based on anisotropic charge transport and related phenomena. For sake of nonvolatile devices, the discovery of advanced ferroelectrics with ferroelectrically switchable anisotropic conductivity and ferroelectric-reversal induced conductivity change (being sizable) is of particular interest.\\

\noindent
\textit{Acknowledgements.} We acknowledge the support from the National Natural Science Foundation of China
(Grants Nos.~12274174, T2225013, 52288102, 52090024, and 12034009). L.B. thanks the Vannevar Bush Faculty Fellowship (VBFF) grant No. N00014-20-1-2834 from the Department of Defense and award No. DMR-1906383 from the National Science Foundation AMASE-i Program (MonArk NSF Quantum Foundry). H.J.Z. thanks the supports from ``Xiaomi YoungScholar'' Project, Arkansas High Performance Computing Center, and high-performance computing center of Jilin University. The authors are grateful to Prof. Zheng Wen at Qingdao University for the valuable discussion.


%\bibliography{ref}

\begin{thebibliography}{72}%
\makeatletter
\providecommand \@ifxundefined [1]{%
 \@ifx{#1\undefined}
}%
\providecommand \@ifnum [1]{%
 \ifnum #1\expandafter \@firstoftwo
 \else \expandafter \@secondoftwo
 \fi
}%
\providecommand \@ifx [1]{%
 \ifx #1\expandafter \@firstoftwo
 \else \expandafter \@secondoftwo
 \fi
}%
\providecommand \natexlab [1]{#1}%
\providecommand \enquote  [1]{``#1''}%
\providecommand \bibnamefont  [1]{#1}%
\providecommand \bibfnamefont [1]{#1}%
\providecommand \citenamefont [1]{#1}%
\providecommand \href@noop [0]{\@secondoftwo}%
\providecommand \href [0]{\begingroup \@sanitize@url \@href}%
\providecommand \@href[1]{\@@startlink{#1}\@@href}%
\providecommand \@@href[1]{\endgroup#1\@@endlink}%
\providecommand \@sanitize@url [0]{\catcode `\\12\catcode `\$12\catcode `\&12\catcode `\#12\catcode `\^12\catcode `\_12\catcode `\%12\relax}%
\providecommand \@@startlink[1]{}%
\providecommand \@@endlink[0]{}%
\providecommand \url  [0]{\begingroup\@sanitize@url \@url }%
\providecommand \@url [1]{\endgroup\@href {#1}{\urlprefix }}%
\providecommand \urlprefix  [0]{URL }%
\providecommand \Eprint [0]{\href }%
\providecommand \doibase [0]{https://doi.org/}%
\providecommand \selectlanguage [0]{\@gobble}%
\providecommand \bibinfo  [0]{\@secondoftwo}%
\providecommand \bibfield  [0]{\@secondoftwo}%
\providecommand \translation [1]{[#1]}%
\providecommand \BibitemOpen [0]{}%
\providecommand \bibitemStop [0]{}%
\providecommand \bibitemNoStop [0]{.\EOS\space}%
\providecommand \EOS [0]{\spacefactor3000\relax}%
\providecommand \BibitemShut  [1]{\csname bibitem#1\endcsname}%
\let\auto@bib@innerbib\@empty
%</preamble>
\bibitem [{\citenamefont {Cui}\ \emph {et~al.}(2025)\citenamefont {Cui}, \citenamefont {Bai},\ and\ \citenamefont {Delin}}]{anisomagnon}%
  \BibitemOpen
  \bibfield  {author} {\bibinfo {author} {\bibfnamefont {Q.}~\bibnamefont {Cui}}, \bibinfo {author} {\bibfnamefont {X.}~\bibnamefont {Bai}},\ and\ \bibinfo {author} {\bibfnamefont {A.}~\bibnamefont {Delin}},\ }\bibfield  {title} {\bibinfo {title} {Anisotropic magnon transport in van der waals ferromagnetic insulators},\ }\href {https://doi.org/10.1002/adfm.202407469} {\bibfield  {journal} {\bibinfo  {journal} {Adv. Funct. Mater.}\ }\textbf {\bibinfo {volume} {35}},\ \bibinfo {pages} {2407469} (\bibinfo {year} {2025})}\BibitemShut {NoStop}%
\bibitem [{\citenamefont {Qi}\ \emph {et~al.}(2023)\citenamefont {Qi}, \citenamefont {Chen}, \citenamefont {Chen}, \citenamefont {Liu}, \citenamefont {Chen}, \citenamefont {Luo}, \citenamefont {Cui}, \citenamefont {Jia}, \citenamefont {Li}, \citenamefont {Huang}, \citenamefont {Song}, \citenamefont {Han}, \citenamefont {Tong}, \citenamefont {Yu}, \citenamefont {Liu}, \citenamefont {Wu}, \citenamefont {Wu}, \citenamefont {Xiao}, \citenamefont {Shindou}, \citenamefont {Xie},\ and\ \citenamefont {Chen}}]{anisomagnon2}%
  \BibitemOpen
  \bibfield  {author} {\bibinfo {author} {\bibfnamefont {S.}~\bibnamefont {Qi}}, \bibinfo {author} {\bibfnamefont {D.}~\bibnamefont {Chen}}, \bibinfo {author} {\bibfnamefont {K.}~\bibnamefont {Chen}}, \bibinfo {author} {\bibfnamefont {J.}~\bibnamefont {Liu}}, \bibinfo {author} {\bibfnamefont {G.}~\bibnamefont {Chen}}, \bibinfo {author} {\bibfnamefont {B.}~\bibnamefont {Luo}}, \bibinfo {author} {\bibfnamefont {H.}~\bibnamefont {Cui}}, \bibinfo {author} {\bibfnamefont {L.}~\bibnamefont {Jia}}, \bibinfo {author} {\bibfnamefont {J.}~\bibnamefont {Li}}, \bibinfo {author} {\bibfnamefont {M.}~\bibnamefont {Huang}}, \bibinfo {author} {\bibfnamefont {Y.}~\bibnamefont {Song}}, \bibinfo {author} {\bibfnamefont {S.}~\bibnamefont {Han}}, \bibinfo {author} {\bibfnamefont {L.}~\bibnamefont {Tong}}, \bibinfo {author} {\bibfnamefont {P.}~\bibnamefont {Yu}}, \bibinfo {author} {\bibfnamefont {Y.}~\bibnamefont {Liu}}, \bibinfo {author} {\bibfnamefont {H.}~\bibnamefont {Wu}}, \bibinfo {author} {\bibfnamefont {S.}~\bibnamefont
  {Wu}}, \bibinfo {author} {\bibfnamefont {J.}~\bibnamefont {Xiao}}, \bibinfo {author} {\bibfnamefont {R.}~\bibnamefont {Shindou}}, \bibinfo {author} {\bibfnamefont {X.~C.}\ \bibnamefont {Xie}},\ and\ \bibinfo {author} {\bibfnamefont {J.-H.}\ \bibnamefont {Chen}},\ }\bibfield  {title} {\bibinfo {title} {Giant electrically tunable magnon transport anisotropy in a van der waals antiferromagnetic insulator},\ }\href {https://doi.org/10.1038/s41467-023-38172-7} {\bibfield  {journal} {\bibinfo  {journal} {Nat. Commun.}\ }\textbf {\bibinfo {volume} {14}},\ \bibinfo {pages} {2526} (\bibinfo {year} {2023})}\BibitemShut {NoStop}%
\bibitem [{\citenamefont {Luo}\ \emph {et~al.}(2023)\citenamefont {Luo}, \citenamefont {Mao}, \citenamefont {Ding}, \citenamefont {Chiu}, \citenamefont {Ji}, \citenamefont {Watanabe}, \citenamefont {Taniguchi}, \citenamefont {Tung}, \citenamefont {Park}, \citenamefont {Kim}, \citenamefont {Kong},\ and\ \citenamefont {Wilson}}]{anisopolar}%
  \BibitemOpen
  \bibfield  {author} {\bibinfo {author} {\bibfnamefont {Y.}~\bibnamefont {Luo}}, \bibinfo {author} {\bibfnamefont {N.}~\bibnamefont {Mao}}, \bibinfo {author} {\bibfnamefont {D.}~\bibnamefont {Ding}}, \bibinfo {author} {\bibfnamefont {M.-H.}\ \bibnamefont {Chiu}}, \bibinfo {author} {\bibfnamefont {X.}~\bibnamefont {Ji}}, \bibinfo {author} {\bibfnamefont {K.}~\bibnamefont {Watanabe}}, \bibinfo {author} {\bibfnamefont {T.}~\bibnamefont {Taniguchi}}, \bibinfo {author} {\bibfnamefont {V.}~\bibnamefont {Tung}}, \bibinfo {author} {\bibfnamefont {H.}~\bibnamefont {Park}}, \bibinfo {author} {\bibfnamefont {P.}~\bibnamefont {Kim}}, \bibinfo {author} {\bibfnamefont {J.}~\bibnamefont {Kong}},\ and\ \bibinfo {author} {\bibfnamefont {W.~L.}\ \bibnamefont {Wilson}},\ }\bibfield  {title} {\bibinfo {title} {Electrically switchable anisotropic polariton propagation in a ferroelectric van der waals semiconductor},\ }\href {https://doi.org/10.1038/s41565-022-01312-z} {\bibfield  {journal} {\bibinfo  {journal} {Nat.
  Nanotechnol.}\ }\textbf {\bibinfo {volume} {18}},\ \bibinfo {pages} {350} (\bibinfo {year} {2023})}\BibitemShut {NoStop}%
\bibitem [{\citenamefont {Wang}\ \emph {et~al.}(2019{\natexlab{a}})\citenamefont {Wang}, \citenamefont {Chen}, \citenamefont {Zhu}, \citenamefont {Wang}, \citenamefont {Dong}, \citenamefont {Sun}, \citenamefont {Zhang}, \citenamefont {Cao}, \citenamefont {Li}, \citenamefont {Huang}, \citenamefont {Zhang}, \citenamefont {Liu}, \citenamefont {Sun}, \citenamefont {Ye}, \citenamefont {Song}, \citenamefont {Wang}, \citenamefont {Han}, \citenamefont {Yang}, \citenamefont {Guo}, \citenamefont {Qin}, \citenamefont {Xiao}, \citenamefont {Zhang}, \citenamefont {Chen}, \citenamefont {Han},\ and\ \citenamefont {Zhang}}]{eca2}%
  \BibitemOpen
  \bibfield  {author} {\bibinfo {author} {\bibfnamefont {H.}~\bibnamefont {Wang}}, \bibinfo {author} {\bibfnamefont {M.-L.}\ \bibnamefont {Chen}}, \bibinfo {author} {\bibfnamefont {M.}~\bibnamefont {Zhu}}, \bibinfo {author} {\bibfnamefont {Y.}~\bibnamefont {Wang}}, \bibinfo {author} {\bibfnamefont {B.}~\bibnamefont {Dong}}, \bibinfo {author} {\bibfnamefont {X.}~\bibnamefont {Sun}}, \bibinfo {author} {\bibfnamefont {X.}~\bibnamefont {Zhang}}, \bibinfo {author} {\bibfnamefont {S.}~\bibnamefont {Cao}}, \bibinfo {author} {\bibfnamefont {X.}~\bibnamefont {Li}}, \bibinfo {author} {\bibfnamefont {J.}~\bibnamefont {Huang}}, \bibinfo {author} {\bibfnamefont {L.}~\bibnamefont {Zhang}}, \bibinfo {author} {\bibfnamefont {W.}~\bibnamefont {Liu}}, \bibinfo {author} {\bibfnamefont {D.}~\bibnamefont {Sun}}, \bibinfo {author} {\bibfnamefont {Y.}~\bibnamefont {Ye}}, \bibinfo {author} {\bibfnamefont {K.}~\bibnamefont {Song}}, \bibinfo {author} {\bibfnamefont {J.}~\bibnamefont {Wang}}, \bibinfo {author} {\bibfnamefont
  {Y.}~\bibnamefont {Han}}, \bibinfo {author} {\bibfnamefont {T.}~\bibnamefont {Yang}}, \bibinfo {author} {\bibfnamefont {H.}~\bibnamefont {Guo}}, \bibinfo {author} {\bibfnamefont {C.}~\bibnamefont {Qin}}, \bibinfo {author} {\bibfnamefont {L.}~\bibnamefont {Xiao}}, \bibinfo {author} {\bibfnamefont {J.}~\bibnamefont {Zhang}}, \bibinfo {author} {\bibfnamefont {J.}~\bibnamefont {Chen}}, \bibinfo {author} {\bibfnamefont {Z.}~\bibnamefont {Han}},\ and\ \bibinfo {author} {\bibfnamefont {Z.}~\bibnamefont {Zhang}},\ }\bibfield  {title} {\bibinfo {title} {Gate tunable giant anisotropic resistance in ultra-thin {G}a{T}e},\ }\href {https://doi.org/10.1038/s41467-019-10256-3} {\bibfield  {journal} {\bibinfo  {journal} {Nat. Commun.}\ }\textbf {\bibinfo {volume} {10}},\ \bibinfo {pages} {2302} (\bibinfo {year} {2019}{\natexlab{a}})}\BibitemShut {NoStop}%
\bibitem [{\citenamefont {Wadley}\ \emph {et~al.}(2016)\citenamefont {Wadley}, \citenamefont {Howells}, \citenamefont {Železný}, \citenamefont {Andrews}, \citenamefont {Hills}, \citenamefont {Campion}, \citenamefont {Novák}, \citenamefont {Olejník}, \citenamefont {Maccherozzi}, \citenamefont {Dhesi}, \citenamefont {Martin}, \citenamefont {Wagner}, \citenamefont {Wunderlich}, \citenamefont {Freimuth}, \citenamefont {Mokrousov}, \citenamefont {Kuneš}, \citenamefont {Chauhan}, \citenamefont {Grzybowski}, \citenamefont {Rushforth}, \citenamefont {Edmonds}, \citenamefont {Gallagher},\ and\ \citenamefont {Jungwirth}}]{amr1}%
  \BibitemOpen
  \bibfield  {author} {\bibinfo {author} {\bibfnamefont {P.}~\bibnamefont {Wadley}}, \bibinfo {author} {\bibfnamefont {B.}~\bibnamefont {Howells}}, \bibinfo {author} {\bibfnamefont {J.}~\bibnamefont {Železný}}, \bibinfo {author} {\bibfnamefont {C.}~\bibnamefont {Andrews}}, \bibinfo {author} {\bibfnamefont {V.}~\bibnamefont {Hills}}, \bibinfo {author} {\bibfnamefont {R.~P.}\ \bibnamefont {Campion}}, \bibinfo {author} {\bibfnamefont {V.}~\bibnamefont {Novák}}, \bibinfo {author} {\bibfnamefont {K.}~\bibnamefont {Olejník}}, \bibinfo {author} {\bibfnamefont {F.}~\bibnamefont {Maccherozzi}}, \bibinfo {author} {\bibfnamefont {S.~S.}\ \bibnamefont {Dhesi}}, \bibinfo {author} {\bibfnamefont {S.~Y.}\ \bibnamefont {Martin}}, \bibinfo {author} {\bibfnamefont {T.}~\bibnamefont {Wagner}}, \bibinfo {author} {\bibfnamefont {J.}~\bibnamefont {Wunderlich}}, \bibinfo {author} {\bibfnamefont {F.}~\bibnamefont {Freimuth}}, \bibinfo {author} {\bibfnamefont {Y.}~\bibnamefont {Mokrousov}}, \bibinfo {author} {\bibfnamefont
  {J.}~\bibnamefont {Kuneš}}, \bibinfo {author} {\bibfnamefont {J.~S.}\ \bibnamefont {Chauhan}}, \bibinfo {author} {\bibfnamefont {M.~J.}\ \bibnamefont {Grzybowski}}, \bibinfo {author} {\bibfnamefont {A.~W.}\ \bibnamefont {Rushforth}}, \bibinfo {author} {\bibfnamefont {K.~W.}\ \bibnamefont {Edmonds}}, \bibinfo {author} {\bibfnamefont {B.~L.}\ \bibnamefont {Gallagher}},\ and\ \bibinfo {author} {\bibfnamefont {T.}~\bibnamefont {Jungwirth}},\ }\bibfield  {title} {\bibinfo {title} {Electrical switching of an antiferromagnet},\ }\href {https://doi.org/10.1126/science.aab1031} {\bibfield  {journal} {\bibinfo  {journal} {Science}\ }\textbf {\bibinfo {volume} {351}},\ \bibinfo {pages} {587} (\bibinfo {year} {2016})}\BibitemShut {NoStop}%
\bibitem [{\citenamefont {Jungwirth}\ \emph {et~al.}(2016)\citenamefont {Jungwirth}, \citenamefont {Marti}, \citenamefont {Wadley},\ and\ \citenamefont {Wunderlich}}]{afmspintronic2}%
  \BibitemOpen
  \bibfield  {author} {\bibinfo {author} {\bibfnamefont {T.}~\bibnamefont {Jungwirth}}, \bibinfo {author} {\bibfnamefont {X.}~\bibnamefont {Marti}}, \bibinfo {author} {\bibfnamefont {P.}~\bibnamefont {Wadley}},\ and\ \bibinfo {author} {\bibfnamefont {J.}~\bibnamefont {Wunderlich}},\ }\bibfield  {title} {\bibinfo {title} {Antiferromagnetic spintronics},\ }\href {https://doi.org/10.1038/nnano.2016.18} {\bibfield  {journal} {\bibinfo  {journal} {Nat. Nanotechnol.}\ }\textbf {\bibinfo {volume} {11}},\ \bibinfo {pages} {231} (\bibinfo {year} {2016})}\BibitemShut {NoStop}%
\bibitem [{\citenamefont {Cording}\ \emph {et~al.}(2024)\citenamefont {Cording}, \citenamefont {Liu}, \citenamefont {Tan}, \citenamefont {Watanabe}, \citenamefont {Taniguchi}, \citenamefont {Avsar},\ and\ \citenamefont {\"{O}zyilmaz}}]{anisospin}%
  \BibitemOpen
  \bibfield  {author} {\bibinfo {author} {\bibfnamefont {L.}~\bibnamefont {Cording}}, \bibinfo {author} {\bibfnamefont {J.}~\bibnamefont {Liu}}, \bibinfo {author} {\bibfnamefont {J.~Y.}\ \bibnamefont {Tan}}, \bibinfo {author} {\bibfnamefont {K.}~\bibnamefont {Watanabe}}, \bibinfo {author} {\bibfnamefont {T.}~\bibnamefont {Taniguchi}}, \bibinfo {author} {\bibfnamefont {A.}~\bibnamefont {Avsar}},\ and\ \bibinfo {author} {\bibfnamefont {B.}~\bibnamefont {\"{O}zyilmaz}},\ }\bibfield  {title} {\bibinfo {title} {Highly anisotropic spin transport in ultrathin black phosphorus},\ }\href {https://doi.org/10.1038/s41563-023-01779-8} {\bibfield  {journal} {\bibinfo  {journal} {Nat. Mater.}\ }\textbf {\bibinfo {volume} {23}},\ \bibinfo {pages} {479} (\bibinfo {year} {2024})}\BibitemShut {NoStop}%
\bibitem [{\citenamefont {Dai}\ \emph {et~al.}(2022)\citenamefont {Dai}, \citenamefont {Zhao}, \citenamefont {Ma}, \citenamefont {Tang}, \citenamefont {Qiu}, \citenamefont {Liu}, \citenamefont {Yuan},\ and\ \citenamefont {Zhou}}]{amr2}%
  \BibitemOpen
  \bibfield  {author} {\bibinfo {author} {\bibfnamefont {Y.}~\bibnamefont {Dai}}, \bibinfo {author} {\bibfnamefont {Y.~W.}\ \bibnamefont {Zhao}}, \bibinfo {author} {\bibfnamefont {L.}~\bibnamefont {Ma}}, \bibinfo {author} {\bibfnamefont {M.}~\bibnamefont {Tang}}, \bibinfo {author} {\bibfnamefont {X.~P.}\ \bibnamefont {Qiu}}, \bibinfo {author} {\bibfnamefont {Y.}~\bibnamefont {Liu}}, \bibinfo {author} {\bibfnamefont {Z.}~\bibnamefont {Yuan}},\ and\ \bibinfo {author} {\bibfnamefont {S.~M.}\ \bibnamefont {Zhou}},\ }\bibfield  {title} {\bibinfo {title} {Fourfold anisotropic magnetoresistance of {L1\textsubscript{0}} {F}e{P}t due to relaxation time anisotropy},\ }\href {https://doi.org/10.1103/PhysRevLett.128.247202} {\bibfield  {journal} {\bibinfo  {journal} {Phys. Rev. Lett.}\ }\textbf {\bibinfo {volume} {128}},\ \bibinfo {pages} {247202} (\bibinfo {year} {2022})}\BibitemShut {NoStop}%
\bibitem [{\citenamefont {Zeng}\ \emph {et~al.}(2020)\citenamefont {Zeng}, \citenamefont {Ren}, \citenamefont {Li}, \citenamefont {Zeng}, \citenamefont {Jia}, \citenamefont {Miao}, \citenamefont {Hoffmann}, \citenamefont {Zhang}, \citenamefont {Wu},\ and\ \citenamefont {Yuan}}]{amr3}%
  \BibitemOpen
  \bibfield  {author} {\bibinfo {author} {\bibfnamefont {F.~L.}\ \bibnamefont {Zeng}}, \bibinfo {author} {\bibfnamefont {Z.~Y.}\ \bibnamefont {Ren}}, \bibinfo {author} {\bibfnamefont {Y.}~\bibnamefont {Li}}, \bibinfo {author} {\bibfnamefont {J.~Y.}\ \bibnamefont {Zeng}}, \bibinfo {author} {\bibfnamefont {M.~W.}\ \bibnamefont {Jia}}, \bibinfo {author} {\bibfnamefont {J.}~\bibnamefont {Miao}}, \bibinfo {author} {\bibfnamefont {A.}~\bibnamefont {Hoffmann}}, \bibinfo {author} {\bibfnamefont {W.}~\bibnamefont {Zhang}}, \bibinfo {author} {\bibfnamefont {Y.~Z.}\ \bibnamefont {Wu}},\ and\ \bibinfo {author} {\bibfnamefont {Z.}~\bibnamefont {Yuan}},\ }\bibfield  {title} {\bibinfo {title} {Intrinsic mechanism for anisotropic magnetoresistance and experimental confirmation in {C}o\textsubscript{\textit{x}}{F}e\textsubscript{1-\textit{x}} single-crystal films},\ }\href {https://doi.org/10.1103/PhysRevLett.125.097201} {\bibfield  {journal} {\bibinfo  {journal} {Phys. Rev. Lett.}\ }\textbf {\bibinfo {volume} {125}},\
  \bibinfo {pages} {097201} (\bibinfo {year} {2020})}\BibitemShut {NoStop}%
\bibitem [{\citenamefont {Wang}\ \emph {et~al.}(2019{\natexlab{b}})\citenamefont {Wang}, \citenamefont {Lu}, \citenamefont {Chen}, \citenamefont {Liu}, \citenamefont {Yuan}, \citenamefont {Cheong}, \citenamefont {Dong},\ and\ \citenamefont {Liu}}]{amr5}%
  \BibitemOpen
  \bibfield  {author} {\bibinfo {author} {\bibfnamefont {H.}~\bibnamefont {Wang}}, \bibinfo {author} {\bibfnamefont {C.}~\bibnamefont {Lu}}, \bibinfo {author} {\bibfnamefont {J.}~\bibnamefont {Chen}}, \bibinfo {author} {\bibfnamefont {Y.}~\bibnamefont {Liu}}, \bibinfo {author} {\bibfnamefont {S.~L.}\ \bibnamefont {Yuan}}, \bibinfo {author} {\bibfnamefont {S.-W.}\ \bibnamefont {Cheong}}, \bibinfo {author} {\bibfnamefont {S.}~\bibnamefont {Dong}},\ and\ \bibinfo {author} {\bibfnamefont {J.-M.}\ \bibnamefont {Liu}},\ }\bibfield  {title} {\bibinfo {title} {Giant anisotropic magnetoresistance and nonvolatile memory in canted antiferromagnet {S}r\textsubscript{2}{I}r{O}\textsubscript{4}},\ }\href {https://doi.org/10.1038/s41467-019-10299-6} {\bibfield  {journal} {\bibinfo  {journal} {Nat. Commun.}\ }\textbf {\bibinfo {volume} {10}},\ \bibinfo {pages} {2280} (\bibinfo {year} {2019}{\natexlab{b}})}\BibitemShut {NoStop}%
\bibitem [{\citenamefont {Liu}\ \emph {et~al.}(2015)\citenamefont {Liu}, \citenamefont {Fu}, \citenamefont {Wang}, \citenamefont {Feng}, \citenamefont {Liu}, \citenamefont {Wan}, \citenamefont {Zhou}, \citenamefont {Wang}, \citenamefont {Shao}, \citenamefont {Ho}, \citenamefont {Huang}, \citenamefont {Cao}, \citenamefont {Wang}, \citenamefont {Li}, \citenamefont {Zeng}, \citenamefont {Song}, \citenamefont {Wang}, \citenamefont {Shi}, \citenamefont {Yuan}, \citenamefont {Hwang}, \citenamefont {Cui}, \citenamefont {Miao},\ and\ \citenamefont {Xing}}]{eca1}%
  \BibitemOpen
  \bibfield  {author} {\bibinfo {author} {\bibfnamefont {E.}~\bibnamefont {Liu}}, \bibinfo {author} {\bibfnamefont {Y.}~\bibnamefont {Fu}}, \bibinfo {author} {\bibfnamefont {Y.}~\bibnamefont {Wang}}, \bibinfo {author} {\bibfnamefont {Y.}~\bibnamefont {Feng}}, \bibinfo {author} {\bibfnamefont {H.}~\bibnamefont {Liu}}, \bibinfo {author} {\bibfnamefont {X.}~\bibnamefont {Wan}}, \bibinfo {author} {\bibfnamefont {W.}~\bibnamefont {Zhou}}, \bibinfo {author} {\bibfnamefont {B.}~\bibnamefont {Wang}}, \bibinfo {author} {\bibfnamefont {L.}~\bibnamefont {Shao}}, \bibinfo {author} {\bibfnamefont {C.-H.}\ \bibnamefont {Ho}}, \bibinfo {author} {\bibfnamefont {Y.-S.}\ \bibnamefont {Huang}}, \bibinfo {author} {\bibfnamefont {Z.}~\bibnamefont {Cao}}, \bibinfo {author} {\bibfnamefont {L.}~\bibnamefont {Wang}}, \bibinfo {author} {\bibfnamefont {A.}~\bibnamefont {Li}}, \bibinfo {author} {\bibfnamefont {J.}~\bibnamefont {Zeng}}, \bibinfo {author} {\bibfnamefont {F.}~\bibnamefont {Song}}, \bibinfo {author} {\bibfnamefont
  {X.}~\bibnamefont {Wang}}, \bibinfo {author} {\bibfnamefont {Y.}~\bibnamefont {Shi}}, \bibinfo {author} {\bibfnamefont {H.}~\bibnamefont {Yuan}}, \bibinfo {author} {\bibfnamefont {H.~Y.}\ \bibnamefont {Hwang}}, \bibinfo {author} {\bibfnamefont {Y.}~\bibnamefont {Cui}}, \bibinfo {author} {\bibfnamefont {F.}~\bibnamefont {Miao}},\ and\ \bibinfo {author} {\bibfnamefont {D.}~\bibnamefont {Xing}},\ }\bibfield  {title} {\bibinfo {title} {Integrated digital inverters based on two-dimensional anisotropic {R}e{S}\textsubscript{2} field-effect transistors},\ }\href {https://doi.org/10.1038/ncomms7991} {\bibfield  {journal} {\bibinfo  {journal} {Nat. Commun.}\ }\textbf {\bibinfo {volume} {6}},\ \bibinfo {pages} {6991} (\bibinfo {year} {2015})}\BibitemShut {NoStop}%
\bibitem [{\citenamefont {Xia}\ \emph {et~al.}(2014)\citenamefont {Xia}, \citenamefont {Wang},\ and\ \citenamefont {Jia}}]{eca3}%
  \BibitemOpen
  \bibfield  {author} {\bibinfo {author} {\bibfnamefont {F.}~\bibnamefont {Xia}}, \bibinfo {author} {\bibfnamefont {H.}~\bibnamefont {Wang}},\ and\ \bibinfo {author} {\bibfnamefont {Y.}~\bibnamefont {Jia}},\ }\bibfield  {title} {\bibinfo {title} {Rediscovering black phosphorus as an anisotropic layered material for optoelectronics and electronics},\ }\href {https://doi.org/10.1038/ncomms5458} {\bibfield  {journal} {\bibinfo  {journal} {Nat. Commun.}\ }\textbf {\bibinfo {volume} {5}},\ \bibinfo {pages} {4458} (\bibinfo {year} {2014})}\BibitemShut {NoStop}%
\bibitem [{\citenamefont {Benítez}\ \emph {et~al.}(2018)\citenamefont {Benítez}, \citenamefont {Sierra}, \citenamefont {Savero~Torres}, \citenamefont {Arrighi}, \citenamefont {Bonell}, \citenamefont {Costache},\ and\ \citenamefont {Valenzuela}}]{anisospin2}%
  \BibitemOpen
  \bibfield  {author} {\bibinfo {author} {\bibfnamefont {L.~A.}\ \bibnamefont {Benítez}}, \bibinfo {author} {\bibfnamefont {J.~F.}\ \bibnamefont {Sierra}}, \bibinfo {author} {\bibfnamefont {W.}~\bibnamefont {Savero~Torres}}, \bibinfo {author} {\bibfnamefont {A.}~\bibnamefont {Arrighi}}, \bibinfo {author} {\bibfnamefont {F.}~\bibnamefont {Bonell}}, \bibinfo {author} {\bibfnamefont {M.~V.}\ \bibnamefont {Costache}},\ and\ \bibinfo {author} {\bibfnamefont {S.~O.}\ \bibnamefont {Valenzuela}},\ }\bibfield  {title} {\bibinfo {title} {Strongly anisotropic spin relaxation in graphene–transition metal dichalcogenide heterostructures at room temperature},\ }\href {https://doi.org/10.1038/s41567-017-0019-2} {\bibfield  {journal} {\bibinfo  {journal} {Nat. Phys.}\ }\textbf {\bibinfo {volume} {14}},\ \bibinfo {pages} {303} (\bibinfo {year} {2018})}\BibitemShut {NoStop}%
\bibitem [{\citenamefont {Zhao}\ \emph {et~al.}(2020)\citenamefont {Zhao}, \citenamefont {Dong}, \citenamefont {Wang}, \citenamefont {Wang}, \citenamefont {Zhang}, \citenamefont {Han},\ and\ \citenamefont {Zhang}}]{aniso}%
  \BibitemOpen
  \bibfield  {author} {\bibinfo {author} {\bibfnamefont {S.}~\bibnamefont {Zhao}}, \bibinfo {author} {\bibfnamefont {B.}~\bibnamefont {Dong}}, \bibinfo {author} {\bibfnamefont {H.}~\bibnamefont {Wang}}, \bibinfo {author} {\bibfnamefont {H.}~\bibnamefont {Wang}}, \bibinfo {author} {\bibfnamefont {Y.}~\bibnamefont {Zhang}}, \bibinfo {author} {\bibfnamefont {Z.~V.}\ \bibnamefont {Han}},\ and\ \bibinfo {author} {\bibfnamefont {H.}~\bibnamefont {Zhang}},\ }\bibfield  {title} {\bibinfo {title} {In-plane anisotropic electronics based on low-symmetry 2{D} materials: progress and prospects},\ }\href {https://doi.org/10.1039/c9na00623k} {\bibfield  {journal} {\bibinfo  {journal} {Nanoscale Adv.}\ }\textbf {\bibinfo {volume} {2}},\ \bibinfo {pages} {109} (\bibinfo {year} {2020})}\BibitemShut {NoStop}%
\bibitem [{\citenamefont {Baltz}\ \emph {et~al.}(2018)\citenamefont {Baltz}, \citenamefont {Manchon}, \citenamefont {Tsoi}, \citenamefont {Moriyama}, \citenamefont {Ono},\ and\ \citenamefont {Tserkovnyak}}]{afmspintronic}%
  \BibitemOpen
  \bibfield  {author} {\bibinfo {author} {\bibfnamefont {V.}~\bibnamefont {Baltz}}, \bibinfo {author} {\bibfnamefont {A.}~\bibnamefont {Manchon}}, \bibinfo {author} {\bibfnamefont {M.}~\bibnamefont {Tsoi}}, \bibinfo {author} {\bibfnamefont {T.}~\bibnamefont {Moriyama}}, \bibinfo {author} {\bibfnamefont {T.}~\bibnamefont {Ono}},\ and\ \bibinfo {author} {\bibfnamefont {Y.}~\bibnamefont {Tserkovnyak}},\ }\bibfield  {title} {\bibinfo {title} {Antiferromagnetic spintronics},\ }\href {https://doi.org/10.1103/RevModPhys.90.015005} {\bibfield  {journal} {\bibinfo  {journal} {Rev. Mod. Phys.}\ }\textbf {\bibinfo {volume} {90}},\ \bibinfo {pages} {015005} (\bibinfo {year} {2018})}\BibitemShut {NoStop}%
\bibitem [{\citenamefont {Ritzinger}\ and\ \citenamefont {Výborný}(2023)}]{amr4}%
  \BibitemOpen
  \bibfield  {author} {\bibinfo {author} {\bibfnamefont {P.}~\bibnamefont {Ritzinger}}\ and\ \bibinfo {author} {\bibfnamefont {K.}~\bibnamefont {Výborný}},\ }\bibfield  {title} {\bibinfo {title} {Anisotropic magnetoresistance: materials, models and applications},\ }\href {https://doi.org/10.1098/rsos.230564} {\bibfield  {journal} {\bibinfo  {journal} {R. Soc. Open Sci.}\ }\textbf {\bibinfo {volume} {10}},\ \bibinfo {pages} {230564} (\bibinfo {year} {2023})}\BibitemShut {NoStop}%
\bibitem [{\citenamefont {Walter}\ \emph {et~al.}(2014)\citenamefont {Walter}, \citenamefont {Viret}, \citenamefont {Singh},\ and\ \citenamefont {Bellaiche}}]{galvanomagnetic}%
  \BibitemOpen
  \bibfield  {author} {\bibinfo {author} {\bibfnamefont {R.}~\bibnamefont {Walter}}, \bibinfo {author} {\bibfnamefont {M.}~\bibnamefont {Viret}}, \bibinfo {author} {\bibfnamefont {S.}~\bibnamefont {Singh}},\ and\ \bibinfo {author} {\bibfnamefont {L.}~\bibnamefont {Bellaiche}},\ }\bibfield  {title} {\bibinfo {title} {Revisiting galvanomagnetic effects in conducting ferromagnets},\ }\href {https://doi.org/10.1088/0953-8984/26/43/432201} {\bibfield  {journal} {\bibinfo  {journal} {J. Phys.: Condens. Matter}\ }\textbf {\bibinfo {volume} {26}},\ \bibinfo {pages} {432201} (\bibinfo {year} {2014})}\BibitemShut {NoStop}%
\bibitem [{\citenamefont {Jia}\ \emph {et~al.}(2024)\citenamefont {Jia}, \citenamefont {Yang}, \citenamefont {Fang}, \citenamefont {Lu}, \citenamefont {Xie}, \citenamefont {Wei}, \citenamefont {Tian}, \citenamefont {Zhang},\ and\ \citenamefont {Yang}}]{ftj1}%
  \BibitemOpen
  \bibfield  {author} {\bibinfo {author} {\bibfnamefont {Y.}~\bibnamefont {Jia}}, \bibinfo {author} {\bibfnamefont {Q.}~\bibnamefont {Yang}}, \bibinfo {author} {\bibfnamefont {Y.-W.}\ \bibnamefont {Fang}}, \bibinfo {author} {\bibfnamefont {Y.}~\bibnamefont {Lu}}, \bibinfo {author} {\bibfnamefont {M.}~\bibnamefont {Xie}}, \bibinfo {author} {\bibfnamefont {J.}~\bibnamefont {Wei}}, \bibinfo {author} {\bibfnamefont {J.}~\bibnamefont {Tian}}, \bibinfo {author} {\bibfnamefont {L.}~\bibnamefont {Zhang}},\ and\ \bibinfo {author} {\bibfnamefont {R.}~\bibnamefont {Yang}},\ }\bibfield  {title} {\bibinfo {title} {Giant tunnelling electroresistance in atomic-scale ferroelectric tunnel junctions},\ }\href {https://doi.org/10.1038/s41467-024-44927-7} {\bibfield  {journal} {\bibinfo  {journal} {Nat. Commun.}\ }\textbf {\bibinfo {volume} {15}},\ \bibinfo {pages} {693} (\bibinfo {year} {2024})}\BibitemShut {NoStop}%
\bibitem [{\citenamefont {Wu}\ \emph {et~al.}(2020)\citenamefont {Wu}, \citenamefont {Chen}, \citenamefont {Yang}, \citenamefont {Cao}, \citenamefont {Yan}, \citenamefont {Liu}, \citenamefont {Sun}, \citenamefont {Ling}, \citenamefont {Guo},\ and\ \citenamefont {Wang}}]{ftj2}%
  \BibitemOpen
  \bibfield  {author} {\bibinfo {author} {\bibfnamefont {J.}~\bibnamefont {Wu}}, \bibinfo {author} {\bibfnamefont {H.-Y.}\ \bibnamefont {Chen}}, \bibinfo {author} {\bibfnamefont {N.}~\bibnamefont {Yang}}, \bibinfo {author} {\bibfnamefont {J.}~\bibnamefont {Cao}}, \bibinfo {author} {\bibfnamefont {X.}~\bibnamefont {Yan}}, \bibinfo {author} {\bibfnamefont {F.}~\bibnamefont {Liu}}, \bibinfo {author} {\bibfnamefont {Q.}~\bibnamefont {Sun}}, \bibinfo {author} {\bibfnamefont {X.}~\bibnamefont {Ling}}, \bibinfo {author} {\bibfnamefont {J.}~\bibnamefont {Guo}},\ and\ \bibinfo {author} {\bibfnamefont {H.}~\bibnamefont {Wang}},\ }\bibfield  {title} {\bibinfo {title} {High tunnelling electroresistance in a ferroelectric van der waals heterojunction via giant barrier height modulation},\ }\href {https://doi.org/10.1038/s41928-020-0441-9} {\bibfield  {journal} {\bibinfo  {journal} {Nat. Electron.}\ }\textbf {\bibinfo {volume} {3}},\ \bibinfo {pages} {466} (\bibinfo {year} {2020})}\BibitemShut {NoStop}%
\bibitem [{\citenamefont {Xi}\ \emph {et~al.}(2017)\citenamefont {Xi}, \citenamefont {Ruan}, \citenamefont {Li}, \citenamefont {Zheng}, \citenamefont {Wen}, \citenamefont {Dai}, \citenamefont {Li},\ and\ \citenamefont {Wu}}]{ftj3}%
  \BibitemOpen
  \bibfield  {author} {\bibinfo {author} {\bibfnamefont {Z.}~\bibnamefont {Xi}}, \bibinfo {author} {\bibfnamefont {J.}~\bibnamefont {Ruan}}, \bibinfo {author} {\bibfnamefont {C.}~\bibnamefont {Li}}, \bibinfo {author} {\bibfnamefont {C.}~\bibnamefont {Zheng}}, \bibinfo {author} {\bibfnamefont {Z.}~\bibnamefont {Wen}}, \bibinfo {author} {\bibfnamefont {J.}~\bibnamefont {Dai}}, \bibinfo {author} {\bibfnamefont {A.}~\bibnamefont {Li}},\ and\ \bibinfo {author} {\bibfnamefont {D.}~\bibnamefont {Wu}},\ }\bibfield  {title} {\bibinfo {title} {Giant tunnelling electroresistance in metal/ferroelectric/semiconductor tunnel junctions by engineering the schottky barrier},\ }\href {https://doi.org/10.1038/ncomms15217} {\bibfield  {journal} {\bibinfo  {journal} {Nat. Commun.}\ }\textbf {\bibinfo {volume} {8}},\ \bibinfo {pages} {15217} (\bibinfo {year} {2017})}\BibitemShut {NoStop}%
\bibitem [{\citenamefont {Wen}\ \emph {et~al.}(2013)\citenamefont {Wen}, \citenamefont {Li}, \citenamefont {Wu}, \citenamefont {Li},\ and\ \citenamefont {Ming}}]{ftj4}%
  \BibitemOpen
  \bibfield  {author} {\bibinfo {author} {\bibfnamefont {Z.}~\bibnamefont {Wen}}, \bibinfo {author} {\bibfnamefont {C.}~\bibnamefont {Li}}, \bibinfo {author} {\bibfnamefont {D.}~\bibnamefont {Wu}}, \bibinfo {author} {\bibfnamefont {A.}~\bibnamefont {Li}},\ and\ \bibinfo {author} {\bibfnamefont {N.}~\bibnamefont {Ming}},\ }\bibfield  {title} {\bibinfo {title} {Ferroelectric-field-effect-enhanced electroresistance in metal/ferroelectric/semiconductor tunnel junctions},\ }\href {https://doi.org/10.1038/nmat3649} {\bibfield  {journal} {\bibinfo  {journal} {Nat. Mater.}\ }\textbf {\bibinfo {volume} {12}},\ \bibinfo {pages} {617–621} (\bibinfo {year} {2013})}\BibitemShut {NoStop}%
\bibitem [{\citenamefont {Zhuravlev}\ \emph {et~al.}(2005)\citenamefont {Zhuravlev}, \citenamefont {Sabirianov}, \citenamefont {Jaswal},\ and\ \citenamefont {Tsymbal}}]{ftj5}%
  \BibitemOpen
  \bibfield  {author} {\bibinfo {author} {\bibfnamefont {M.~Y.}\ \bibnamefont {Zhuravlev}}, \bibinfo {author} {\bibfnamefont {R.~F.}\ \bibnamefont {Sabirianov}}, \bibinfo {author} {\bibfnamefont {S.~S.}\ \bibnamefont {Jaswal}},\ and\ \bibinfo {author} {\bibfnamefont {E.~Y.}\ \bibnamefont {Tsymbal}},\ }\bibfield  {title} {\bibinfo {title} {Giant electroresistance in ferroelectric tunnel junctions},\ }\href {https://doi.org/10.1103/PhysRevLett.94.246802} {\bibfield  {journal} {\bibinfo  {journal} {Phys. Rev. Lett.}\ }\textbf {\bibinfo {volume} {94}},\ \bibinfo {pages} {246802} (\bibinfo {year} {2005})}\BibitemShut {NoStop}%
\bibitem [{\citenamefont {Ding}\ \emph {et~al.}(2021)\citenamefont {Ding}, \citenamefont {Shao}, \citenamefont {Li}, \citenamefont {Wen},\ and\ \citenamefont {Tsymbal}}]{ftj6}%
  \BibitemOpen
  \bibfield  {author} {\bibinfo {author} {\bibfnamefont {J.}~\bibnamefont {Ding}}, \bibinfo {author} {\bibfnamefont {D.-F.}\ \bibnamefont {Shao}}, \bibinfo {author} {\bibfnamefont {M.}~\bibnamefont {Li}}, \bibinfo {author} {\bibfnamefont {L.-W.}\ \bibnamefont {Wen}},\ and\ \bibinfo {author} {\bibfnamefont {E.~Y.}\ \bibnamefont {Tsymbal}},\ }\bibfield  {title} {\bibinfo {title} {Two-dimensional antiferroelectric tunnel junction},\ }\href {https://doi.org/10.1103/PhysRevLett.126.057601} {\bibfield  {journal} {\bibinfo  {journal} {Phys. Rev. Lett.}\ }\textbf {\bibinfo {volume} {126}},\ \bibinfo {pages} {057601} (\bibinfo {year} {2021})}\BibitemShut {NoStop}%
\bibitem [{\citenamefont {Liu}\ \emph {et~al.}(2016)\citenamefont {Liu}, \citenamefont {Burton},\ and\ \citenamefont {Tsymbal}}]{ftj7}%
  \BibitemOpen
  \bibfield  {author} {\bibinfo {author} {\bibfnamefont {X.}~\bibnamefont {Liu}}, \bibinfo {author} {\bibfnamefont {J.~D.}\ \bibnamefont {Burton}},\ and\ \bibinfo {author} {\bibfnamefont {E.~Y.}\ \bibnamefont {Tsymbal}},\ }\bibfield  {title} {\bibinfo {title} {Enhanced tunneling electroresistance in ferroelectric tunnel junctions due to the reversible metallization of the barrier},\ }\href {https://doi.org/10.1103/PhysRevLett.116.197602} {\bibfield  {journal} {\bibinfo  {journal} {Phys. Rev. Lett.}\ }\textbf {\bibinfo {volume} {116}},\ \bibinfo {pages} {197602} (\bibinfo {year} {2016})}\BibitemShut {NoStop}%
\bibitem [{\citenamefont {Hernandez-Martin}\ \emph {et~al.}(2020)\citenamefont {Hernandez-Martin}, \citenamefont {Gallego}, \citenamefont {Tornos}, \citenamefont {Rouco}, \citenamefont {Beltran}, \citenamefont {Munuera}, \citenamefont {Sanchez-Manzano}, \citenamefont {Cabero}, \citenamefont {Cuellar}, \citenamefont {Arias}, \citenamefont {Sanchez-Santolino}, \citenamefont {Mompean}, \citenamefont {Garcia-Hernandez}, \citenamefont {Rivera-Calzada}, \citenamefont {Pennycook}, \citenamefont {Varela}, \citenamefont {Mu\~noz}, \citenamefont {Sefrioui}, \citenamefont {Leon},\ and\ \citenamefont {Santamaria}}]{ftj8}%
  \BibitemOpen
  \bibfield  {author} {\bibinfo {author} {\bibfnamefont {D.}~\bibnamefont {Hernandez-Martin}}, \bibinfo {author} {\bibfnamefont {F.}~\bibnamefont {Gallego}}, \bibinfo {author} {\bibfnamefont {J.}~\bibnamefont {Tornos}}, \bibinfo {author} {\bibfnamefont {V.}~\bibnamefont {Rouco}}, \bibinfo {author} {\bibfnamefont {J.~I.}\ \bibnamefont {Beltran}}, \bibinfo {author} {\bibfnamefont {C.}~\bibnamefont {Munuera}}, \bibinfo {author} {\bibfnamefont {D.}~\bibnamefont {Sanchez-Manzano}}, \bibinfo {author} {\bibfnamefont {M.}~\bibnamefont {Cabero}}, \bibinfo {author} {\bibfnamefont {F.}~\bibnamefont {Cuellar}}, \bibinfo {author} {\bibfnamefont {D.}~\bibnamefont {Arias}}, \bibinfo {author} {\bibfnamefont {G.}~\bibnamefont {Sanchez-Santolino}}, \bibinfo {author} {\bibfnamefont {F.~J.}\ \bibnamefont {Mompean}}, \bibinfo {author} {\bibfnamefont {M.}~\bibnamefont {Garcia-Hernandez}}, \bibinfo {author} {\bibfnamefont {A.}~\bibnamefont {Rivera-Calzada}}, \bibinfo {author} {\bibfnamefont {S.~J.}\ \bibnamefont {Pennycook}},
  \bibinfo {author} {\bibfnamefont {M.}~\bibnamefont {Varela}}, \bibinfo {author} {\bibfnamefont {M.~C.}\ \bibnamefont {Mu\~noz}}, \bibinfo {author} {\bibfnamefont {Z.}~\bibnamefont {Sefrioui}}, \bibinfo {author} {\bibfnamefont {C.}~\bibnamefont {Leon}},\ and\ \bibinfo {author} {\bibfnamefont {J.}~\bibnamefont {Santamaria}},\ }\bibfield  {title} {\bibinfo {title} {Controlled sign reversal of electroresistance in oxide tunnel junctions by electrochemical-ferroelectric coupling},\ }\href {https://doi.org/10.1103/PhysRevLett.125.266802} {\bibfield  {journal} {\bibinfo  {journal} {Phys. Rev. Lett.}\ }\textbf {\bibinfo {volume} {125}},\ \bibinfo {pages} {266802} (\bibinfo {year} {2020})}\BibitemShut {NoStop}%
\bibitem [{\citenamefont {Klyukin}\ \emph {et~al.}(2018)\citenamefont {Klyukin}, \citenamefont {Tao}, \citenamefont {Tsymbal},\ and\ \citenamefont {Alexandrov}}]{ftj9}%
  \BibitemOpen
  \bibfield  {author} {\bibinfo {author} {\bibfnamefont {K.}~\bibnamefont {Klyukin}}, \bibinfo {author} {\bibfnamefont {L.~L.}\ \bibnamefont {Tao}}, \bibinfo {author} {\bibfnamefont {E.~Y.}\ \bibnamefont {Tsymbal}},\ and\ \bibinfo {author} {\bibfnamefont {V.}~\bibnamefont {Alexandrov}},\ }\bibfield  {title} {\bibinfo {title} {Defect-assisted tunneling electroresistance in ferroelectric tunnel junctions},\ }\href {https://doi.org/10.1103/PhysRevLett.121.056601} {\bibfield  {journal} {\bibinfo  {journal} {Phys. Rev. Lett.}\ }\textbf {\bibinfo {volume} {121}},\ \bibinfo {pages} {056601} (\bibinfo {year} {2018})}\BibitemShut {NoStop}%
\bibitem [{Note1()}]{Note1}%
  \BibitemOpen
  \bibinfo {note} {According to Eq.~(\ref {eq:ecapol}), the $\sigma _{\chi ,\chi }(P) - \sigma _{\chi ,\chi }(-P)$ conductivity difference along $\protect \bm {\chi }$ direction equals $\sigma _{\chi ^\prime ,\chi ^\prime }(P) + \sigma _{\chi ^\prime ,\chi ^\prime }(-P) - 2\sigma _{\chi ,\chi }(-P)$, which is not necessarily zero. The similar logic applies to the conductivity difference along $\protect \bm {\chi ^\prime }$ direction. In next section, we shall discuss this point in details.}\BibitemShut {Stop}%
\bibitem [{\citenamefont {Aroyo}\ \emph {et~al.}(2006{\natexlab{a}})\citenamefont {Aroyo}, \citenamefont {Kirov}, \citenamefont {Capillas}, \citenamefont {Perez-Mato},\ and\ \citenamefont {Wondratschek}}]{aroyo2006}%
  \BibitemOpen
  \bibfield  {author} {\bibinfo {author} {\bibfnamefont {M.~I.}\ \bibnamefont {Aroyo}}, \bibinfo {author} {\bibfnamefont {A.}~\bibnamefont {Kirov}}, \bibinfo {author} {\bibfnamefont {C.}~\bibnamefont {Capillas}}, \bibinfo {author} {\bibfnamefont {J.~M.}\ \bibnamefont {Perez-Mato}},\ and\ \bibinfo {author} {\bibfnamefont {H.}~\bibnamefont {Wondratschek}},\ }\bibfield  {title} {\bibinfo {title} {Bilbao crystallographic server. {II}. representations of crystallographic point groups and space groups},\ }\href {https://doi.org/10.1107/s0108767305040286} {\bibfield  {journal} {\bibinfo  {journal} {Acta Cryst. A}\ }\textbf {\bibinfo {volume} {62}},\ \bibinfo {pages} {115} (\bibinfo {year} {2006}{\natexlab{a}})}\BibitemShut {NoStop}%
\bibitem [{\citenamefont {Aroyo}\ \emph {et~al.}(2006{\natexlab{b}})\citenamefont {Aroyo}, \citenamefont {Perez-Mato}, \citenamefont {Capillas}, \citenamefont {Kroumova}, \citenamefont {Ivantchev}, \citenamefont {Madariaga}, \citenamefont {Kirov},\ and\ \citenamefont {Wondratschek}}]{aroyo20062}%
  \BibitemOpen
  \bibfield  {author} {\bibinfo {author} {\bibfnamefont {M.~I.}\ \bibnamefont {Aroyo}}, \bibinfo {author} {\bibfnamefont {J.~M.}\ \bibnamefont {Perez-Mato}}, \bibinfo {author} {\bibfnamefont {C.}~\bibnamefont {Capillas}}, \bibinfo {author} {\bibfnamefont {E.}~\bibnamefont {Kroumova}}, \bibinfo {author} {\bibfnamefont {S.}~\bibnamefont {Ivantchev}}, \bibinfo {author} {\bibfnamefont {G.}~\bibnamefont {Madariaga}}, \bibinfo {author} {\bibfnamefont {A.}~\bibnamefont {Kirov}},\ and\ \bibinfo {author} {\bibfnamefont {H.}~\bibnamefont {Wondratschek}},\ }\bibfield  {title} {\bibinfo {title} {Bilbao crystallographic server: I. databases and crystallographic computing programs},\ }\href {https://doi.org/10.1524/zkri.2006.221.1.15} {\bibfield  {journal} {\bibinfo  {journal} {Z. Kristallogr. Cryst. Mater.}\ }\textbf {\bibinfo {volume} {221}},\ \bibinfo {pages} {15} (\bibinfo {year} {2006}{\natexlab{b}})}\BibitemShut {NoStop}%
\bibitem [{\citenamefont {Aroyo}\ \emph {et~al.}(2011)\citenamefont {Aroyo}, \citenamefont {Perez-Mato}, \citenamefont {Orobengoa}, \citenamefont {Tasci}, \citenamefont {de~la Flor},\ and\ \citenamefont {Kirov}}]{aroyo2011}%
  \BibitemOpen
  \bibfield  {author} {\bibinfo {author} {\bibfnamefont {M.~I.}\ \bibnamefont {Aroyo}}, \bibinfo {author} {\bibfnamefont {J.~M.}\ \bibnamefont {Perez-Mato}}, \bibinfo {author} {\bibfnamefont {D.}~\bibnamefont {Orobengoa}}, \bibinfo {author} {\bibfnamefont {E.}~\bibnamefont {Tasci}}, \bibinfo {author} {\bibfnamefont {G.}~\bibnamefont {de~la Flor}},\ and\ \bibinfo {author} {\bibfnamefont {A.}~\bibnamefont {Kirov}},\ }\bibfield  {title} {\bibinfo {title} {Crystallography online: Bilbao crystallographic server},\ }\href@noop {} {\bibfield  {journal} {\bibinfo  {journal} {Bulg. Chem. Commun.}\ }\textbf {\bibinfo {volume} {43}},\ \bibinfo {pages} {183} (\bibinfo {year} {2011})}\BibitemShut {NoStop}%
\bibitem [{\citenamefont {Hinuma}\ \emph {et~al.}(2017)\citenamefont {Hinuma}, \citenamefont {Pizzi}, \citenamefont {Kumagai}, \citenamefont {Oba},\ and\ \citenamefont {Tanaka}}]{seekpath}%
  \BibitemOpen
  \bibfield  {author} {\bibinfo {author} {\bibfnamefont {Y.}~\bibnamefont {Hinuma}}, \bibinfo {author} {\bibfnamefont {G.}~\bibnamefont {Pizzi}}, \bibinfo {author} {\bibfnamefont {Y.}~\bibnamefont {Kumagai}}, \bibinfo {author} {\bibfnamefont {F.}~\bibnamefont {Oba}},\ and\ \bibinfo {author} {\bibfnamefont {I.}~\bibnamefont {Tanaka}},\ }\bibfield  {title} {\bibinfo {title} {Band structure diagram paths based on crystallography},\ }\href {https://doi.org/10.1016/j.commatsci.2016.10.015} {\bibfield  {journal} {\bibinfo  {journal} {Comp. Mater. Sci.}\ }\textbf {\bibinfo {volume} {128}},\ \bibinfo {pages} {140} (\bibinfo {year} {2017})}\BibitemShut {NoStop}%
\bibitem [{\citenamefont {Togo}\ and\ \citenamefont {Tanaka}(2018)}]{seekpath2}%
  \BibitemOpen
  \bibfield  {author} {\bibinfo {author} {\bibfnamefont {A.}~\bibnamefont {Togo}}\ and\ \bibinfo {author} {\bibfnamefont {I.}~\bibnamefont {Tanaka}},\ }\href@noop {} {\bibinfo {title} {$\texttt{Spglib}$: a software library for crystal symmetry search}} (\bibinfo {year} {2018}),\ \Eprint {https://arxiv.org/abs/arXiv:1808.01590} {arXiv:1808.01590} \BibitemShut {NoStop}%
\bibitem [{\citenamefont {Inc.}()}]{mma}%
  \BibitemOpen
  \bibfield  {author} {\bibinfo {author} {\bibfnamefont {W.~R.}\ \bibnamefont {Inc.}},\ }\href {https://www.wolfram.com/mathematica} {\bibinfo {title} {Mathematica, {V}ersion 12.0}},\ \bibinfo {note} {champaign, IL, 2019}\BibitemShut {NoStop}%
\bibitem [{\citenamefont {Hunter}(2007)}]{hunter2007matplotlib}%
  \BibitemOpen
  \bibfield  {author} {\bibinfo {author} {\bibfnamefont {J.~D.}\ \bibnamefont {Hunter}},\ }\bibfield  {title} {\bibinfo {title} {Matplotlib: A 2{D} graphics environment},\ }\href@noop {} {\bibfield  {journal} {\bibinfo  {journal} {Comput. Sci. Eng.}\ }\textbf {\bibinfo {volume} {9}},\ \bibinfo {pages} {90} (\bibinfo {year} {2007})}\BibitemShut {NoStop}%
\bibitem [{\citenamefont {Ong}\ \emph {et~al.}(2013)\citenamefont {Ong}, \citenamefont {Richards}, \citenamefont {Jain}, \citenamefont {Hautier}, \citenamefont {Kocher}, \citenamefont {Cholia}, \citenamefont {Gunter}, \citenamefont {Chevrier}, \citenamefont {Persson},\ and\ \citenamefont {Ceder}}]{pymatgen}%
  \BibitemOpen
  \bibfield  {author} {\bibinfo {author} {\bibfnamefont {S.~P.}\ \bibnamefont {Ong}}, \bibinfo {author} {\bibfnamefont {W.~D.}\ \bibnamefont {Richards}}, \bibinfo {author} {\bibfnamefont {A.}~\bibnamefont {Jain}}, \bibinfo {author} {\bibfnamefont {G.}~\bibnamefont {Hautier}}, \bibinfo {author} {\bibfnamefont {M.}~\bibnamefont {Kocher}}, \bibinfo {author} {\bibfnamefont {S.}~\bibnamefont {Cholia}}, \bibinfo {author} {\bibfnamefont {D.}~\bibnamefont {Gunter}}, \bibinfo {author} {\bibfnamefont {V.~L.}\ \bibnamefont {Chevrier}}, \bibinfo {author} {\bibfnamefont {K.~A.}\ \bibnamefont {Persson}},\ and\ \bibinfo {author} {\bibfnamefont {G.}~\bibnamefont {Ceder}},\ }\bibfield  {title} {\bibinfo {title} {Python materials genomics (pymatgen): A robust, open-source python library for materials analysis},\ }\href {https://doi.org/10.1016/j.commatsci.2012.10.028} {\bibfield  {journal} {\bibinfo  {journal} {Comput. Mater. Sci.}\ }\textbf {\bibinfo {volume} {68}},\ \bibinfo {pages} {314} (\bibinfo {year}
  {2013})}\BibitemShut {NoStop}%
\bibitem [{\citenamefont {Momma}\ and\ \citenamefont {Izumi}(2011)}]{momma2011vesta}%
  \BibitemOpen
  \bibfield  {author} {\bibinfo {author} {\bibfnamefont {K.}~\bibnamefont {Momma}}\ and\ \bibinfo {author} {\bibfnamefont {F.}~\bibnamefont {Izumi}},\ }\bibfield  {title} {\bibinfo {title} {{VESTA} 3 for three-dimensional visualization of crystal, volumetric and morphology data},\ }\href@noop {} {\bibfield  {journal} {\bibinfo  {journal} {J. Appl. Crystallogr.}\ }\textbf {\bibinfo {volume} {44}},\ \bibinfo {pages} {1272} (\bibinfo {year} {2011})}\BibitemShut {NoStop}%
\bibitem [{\citenamefont {Kresse}\ and\ \citenamefont {Furthm{\"u}ller}(1996)}]{kresse1996efficient}%
  \BibitemOpen
  \bibfield  {author} {\bibinfo {author} {\bibfnamefont {G.}~\bibnamefont {Kresse}}\ and\ \bibinfo {author} {\bibfnamefont {J.}~\bibnamefont {Furthm{\"u}ller}},\ }\bibfield  {title} {\bibinfo {title} {Efficient iterative schemes for ab initio total-energy calculations using a plane-wave basis set},\ }\href@noop {} {\bibfield  {journal} {\bibinfo  {journal} {Phys. Rev. B}\ }\textbf {\bibinfo {volume} {54}},\ \bibinfo {pages} {11169} (\bibinfo {year} {1996})}\BibitemShut {NoStop}%
\bibitem [{\citenamefont {Kresse}\ and\ \citenamefont {Joubert}(1999)}]{kresse1999ultrasoft}%
  \BibitemOpen
  \bibfield  {author} {\bibinfo {author} {\bibfnamefont {G.}~\bibnamefont {Kresse}}\ and\ \bibinfo {author} {\bibfnamefont {D.}~\bibnamefont {Joubert}},\ }\bibfield  {title} {\bibinfo {title} {From ultrasoft pseudopotentials to the projector augmented-wave method},\ }\href@noop {} {\bibfield  {journal} {\bibinfo  {journal} {Phys. Rev. B}\ }\textbf {\bibinfo {volume} {59}},\ \bibinfo {pages} {1758} (\bibinfo {year} {1999})}\BibitemShut {NoStop}%
\bibitem [{\citenamefont {Bl{\"o}chl}(1994)}]{blochl1994projector}%
  \BibitemOpen
  \bibfield  {author} {\bibinfo {author} {\bibfnamefont {P.~E.}\ \bibnamefont {Bl{\"o}chl}},\ }\bibfield  {title} {\bibinfo {title} {Projector augmented-wave method},\ }\href@noop {} {\bibfield  {journal} {\bibinfo  {journal} {Phys. Rev. B}\ }\textbf {\bibinfo {volume} {50}},\ \bibinfo {pages} {17953} (\bibinfo {year} {1994})}\BibitemShut {NoStop}%
\bibitem [{\citenamefont {Madsen}\ \emph {et~al.}(2018)\citenamefont {Madsen}, \citenamefont {Carrete},\ and\ \citenamefont {Verstraete}}]{boltztrap2}%
  \BibitemOpen
  \bibfield  {author} {\bibinfo {author} {\bibfnamefont {G.~K.}\ \bibnamefont {Madsen}}, \bibinfo {author} {\bibfnamefont {J.}~\bibnamefont {Carrete}},\ and\ \bibinfo {author} {\bibfnamefont {M.~J.}\ \bibnamefont {Verstraete}},\ }\bibfield  {title} {\bibinfo {title} {Boltztrap2, a program for interpolating band structures and calculating semi-classical transport coefficients},\ }\href {https://doi.org/10.1016/j.cpc.2018.05.010} {\bibfield  {journal} {\bibinfo  {journal} {Comput. Phys. Commun.}\ }\textbf {\bibinfo {volume} {231}},\ \bibinfo {pages} {140} (\bibinfo {year} {2018})}\BibitemShut {NoStop}%
\bibitem [{\citenamefont {Perdew}\ \emph {et~al.}(2008)\citenamefont {Perdew}, \citenamefont {Ruzsinszky}, \citenamefont {Csonka}, \citenamefont {Vydrov}, \citenamefont {Scuseria}, \citenamefont {Constantin}, \citenamefont {Zhou},\ and\ \citenamefont {Burke}}]{pbesol}%
  \BibitemOpen
  \bibfield  {author} {\bibinfo {author} {\bibfnamefont {J.~P.}\ \bibnamefont {Perdew}}, \bibinfo {author} {\bibfnamefont {A.}~\bibnamefont {Ruzsinszky}}, \bibinfo {author} {\bibfnamefont {G.~I.}\ \bibnamefont {Csonka}}, \bibinfo {author} {\bibfnamefont {O.~A.}\ \bibnamefont {Vydrov}}, \bibinfo {author} {\bibfnamefont {G.~E.}\ \bibnamefont {Scuseria}}, \bibinfo {author} {\bibfnamefont {L.~A.}\ \bibnamefont {Constantin}}, \bibinfo {author} {\bibfnamefont {X.}~\bibnamefont {Zhou}},\ and\ \bibinfo {author} {\bibfnamefont {K.}~\bibnamefont {Burke}},\ }\bibfield  {title} {\bibinfo {title} {Restoring the density-gradient expansion for exchange in solids and surfaces},\ }\href@noop {} {\bibfield  {journal} {\bibinfo  {journal} {Phys. Rev. Lett.}\ }\textbf {\bibinfo {volume} {100}},\ \bibinfo {pages} {136406} (\bibinfo {year} {2008})}\BibitemShut {NoStop}%
\bibitem [{Note2()}]{Note2}%
  \BibitemOpen
  \bibinfo {note} {See Supplementary Material at {\protect \color {blue}a link} which contains general remarks on polarization-switchable ECA, symmetry analysis on ECA, and some numerical results.}\BibitemShut {Stop}%
\bibitem [{\citenamefont {Freimuth}\ \emph {et~al.}(2014)\citenamefont {Freimuth}, \citenamefont {Bl\"ugel},\ and\ \citenamefont {Mokrousov}}]{thetafunction}%
  \BibitemOpen
  \bibfield  {author} {\bibinfo {author} {\bibfnamefont {F.}~\bibnamefont {Freimuth}}, \bibinfo {author} {\bibfnamefont {S.}~\bibnamefont {Bl\"ugel}},\ and\ \bibinfo {author} {\bibfnamefont {Y.}~\bibnamefont {Mokrousov}},\ }\bibfield  {title} {\bibinfo {title} {Spin-orbit torques in {C}o/{P}t(111) and {M}n/{W}(001) magnetic bilayers from first principles},\ }\href {https://doi.org/10.1103/PhysRevB.90.174423} {\bibfield  {journal} {\bibinfo  {journal} {Phys. Rev. B}\ }\textbf {\bibinfo {volume} {90}},\ \bibinfo {pages} {174423} (\bibinfo {year} {2014})}\BibitemShut {NoStop}%
\bibitem [{\citenamefont {Dong}\ and\ \citenamefont {DiSalvo}(2007)}]{nasrp2007}%
  \BibitemOpen
  \bibfield  {author} {\bibinfo {author} {\bibfnamefont {Y.}~\bibnamefont {Dong}}\ and\ \bibinfo {author} {\bibfnamefont {F.~J.}\ \bibnamefont {DiSalvo}},\ }\bibfield  {title} {\bibinfo {title} {Synthesis and single crystal structures of ternary phosphides {L}i\textsubscript{4}{S}r{P}\textsubscript{2} and \textit{AAe}p (\textit{A}={L}i, {N}a; \textit{Ae}={S}r, {B}a)},\ }\href {https://doi.org/10.1016/j.jssc.2006.10.033} {\bibfield  {journal} {\bibinfo  {journal} {J. Solid State Chem.}\ }\textbf {\bibinfo {volume} {180}},\ \bibinfo {pages} {432} (\bibinfo {year} {2007})}\BibitemShut {NoStop}%
\bibitem [{\citenamefont {Fu}\ and\ \citenamefont {Bellaiche}(2003)}]{efield}%
  \BibitemOpen
  \bibfield  {author} {\bibinfo {author} {\bibfnamefont {H.}~\bibnamefont {Fu}}\ and\ \bibinfo {author} {\bibfnamefont {L.}~\bibnamefont {Bellaiche}},\ }\bibfield  {title} {\bibinfo {title} {First-principles determination of electromechanical responses of solids under finite electric fields},\ }\href {https://doi.org/10.1103/PhysRevLett.91.057601} {\bibfield  {journal} {\bibinfo  {journal} {Phys. Rev. Lett.}\ }\textbf {\bibinfo {volume} {91}},\ \bibinfo {pages} {057601} (\bibinfo {year} {2003})}\BibitemShut {NoStop}%
\bibitem [{\citenamefont {Chen}\ \emph {et~al.}(2019)\citenamefont {Chen}, \citenamefont {Xu}, \citenamefont {Tian}, \citenamefont {Xiang}, \citenamefont {{\'I}{\~n}iguez}, \citenamefont {Yang},\ and\ \citenamefont {Bellaiche}}]{chen2019electric}%
  \BibitemOpen
  \bibfield  {author} {\bibinfo {author} {\bibfnamefont {L.}~\bibnamefont {Chen}}, \bibinfo {author} {\bibfnamefont {C.}~\bibnamefont {Xu}}, \bibinfo {author} {\bibfnamefont {H.}~\bibnamefont {Tian}}, \bibinfo {author} {\bibfnamefont {H.}~\bibnamefont {Xiang}}, \bibinfo {author} {\bibfnamefont {J.}~\bibnamefont {{\'I}{\~n}iguez}}, \bibinfo {author} {\bibfnamefont {Y.}~\bibnamefont {Yang}},\ and\ \bibinfo {author} {\bibfnamefont {L.}~\bibnamefont {Bellaiche}},\ }\bibfield  {title} {\bibinfo {title} {Electric-field control of magnetization, jahn-teller distortion, and orbital ordering in ferroelectric ferromagnets},\ }\href@noop {} {\bibfield  {journal} {\bibinfo  {journal} {Physical Review Letters}\ }\textbf {\bibinfo {volume} {122}},\ \bibinfo {pages} {247701} (\bibinfo {year} {2019})}\BibitemShut {NoStop}%
\bibitem [{\citenamefont {Chen}\ \emph {et~al.}(2016)\citenamefont {Chen}, \citenamefont {Yang},\ and\ \citenamefont {Meng}}]{chen2016giant}%
  \BibitemOpen
  \bibfield  {author} {\bibinfo {author} {\bibfnamefont {L.}~\bibnamefont {Chen}}, \bibinfo {author} {\bibfnamefont {Y.}~\bibnamefont {Yang}},\ and\ \bibinfo {author} {\bibfnamefont {X.}~\bibnamefont {Meng}},\ }\bibfield  {title} {\bibinfo {title} {Giant electric-field-induced strain in lead-free piezoelectric materials},\ }\href@noop {} {\bibfield  {journal} {\bibinfo  {journal} {Sci. Rep.}\ }\textbf {\bibinfo {volume} {6}},\ \bibinfo {pages} {25346} (\bibinfo {year} {2016})}\BibitemShut {NoStop}%
\bibitem [{\citenamefont {Zhao}\ \emph {et~al.}(2018)\citenamefont {Zhao}, \citenamefont {Filippetti}, \citenamefont {Escorihuela-Sayalero}, \citenamefont {Delugas}, \citenamefont {Canadell}, \citenamefont {Bellaiche}, \citenamefont {Fiorentini},\ and\ \citenamefont {\'I\~niguez}}]{doping}%
  \BibitemOpen
  \bibfield  {author} {\bibinfo {author} {\bibfnamefont {H.~J.}\ \bibnamefont {Zhao}}, \bibinfo {author} {\bibfnamefont {A.}~\bibnamefont {Filippetti}}, \bibinfo {author} {\bibfnamefont {C.}~\bibnamefont {Escorihuela-Sayalero}}, \bibinfo {author} {\bibfnamefont {P.}~\bibnamefont {Delugas}}, \bibinfo {author} {\bibfnamefont {E.}~\bibnamefont {Canadell}}, \bibinfo {author} {\bibfnamefont {L.}~\bibnamefont {Bellaiche}}, \bibinfo {author} {\bibfnamefont {V.}~\bibnamefont {Fiorentini}},\ and\ \bibinfo {author} {\bibfnamefont {J.}~\bibnamefont {\'I\~niguez}},\ }\bibfield  {title} {\bibinfo {title} {Meta-screening and permanence of polar distortion in metallized ferroelectrics},\ }\href {https://doi.org/10.1103/PhysRevB.97.054107} {\bibfield  {journal} {\bibinfo  {journal} {Phys. Rev. B}\ }\textbf {\bibinfo {volume} {97}},\ \bibinfo {pages} {054107} (\bibinfo {year} {2018})}\BibitemShut {NoStop}%
\bibitem [{\citenamefont {Ueno}\ \emph {et~al.}(2011)\citenamefont {Ueno}, \citenamefont {Nakamura}, \citenamefont {Shimotani}, \citenamefont {Yuan}, \citenamefont {Kimura}, \citenamefont {Nojima}, \citenamefont {Aoki}, \citenamefont {Iwasa},\ and\ \citenamefont {Kawasaki}}]{dopconcen}%
  \BibitemOpen
  \bibfield  {author} {\bibinfo {author} {\bibfnamefont {K.}~\bibnamefont {Ueno}}, \bibinfo {author} {\bibfnamefont {S.}~\bibnamefont {Nakamura}}, \bibinfo {author} {\bibfnamefont {H.}~\bibnamefont {Shimotani}}, \bibinfo {author} {\bibfnamefont {H.~T.}\ \bibnamefont {Yuan}}, \bibinfo {author} {\bibfnamefont {N.}~\bibnamefont {Kimura}}, \bibinfo {author} {\bibfnamefont {T.}~\bibnamefont {Nojima}}, \bibinfo {author} {\bibfnamefont {H.}~\bibnamefont {Aoki}}, \bibinfo {author} {\bibfnamefont {Y.}~\bibnamefont {Iwasa}},\ and\ \bibinfo {author} {\bibfnamefont {M.}~\bibnamefont {Kawasaki}},\ }\bibfield  {title} {\bibinfo {title} {Discovery of superconductivity in ktao3 by electrostatic carrier doping},\ }\href {https://doi.org/10.1038/nnano.2011.78} {\bibfield  {journal} {\bibinfo  {journal} {Nat. Nanotech.}\ }\textbf {\bibinfo {volume} {6}},\ \bibinfo {pages} {408} (\bibinfo {year} {2011})}\BibitemShut {NoStop}%
\bibitem [{\citenamefont {Tsirkin}(2021)}]{drude1}%
  \BibitemOpen
  \bibfield  {author} {\bibinfo {author} {\bibfnamefont {S.~S.}\ \bibnamefont {Tsirkin}},\ }\bibfield  {title} {\bibinfo {title} {High performance wannier interpolation of berry curvature and related quantities with wannierberri code},\ }\href {https://doi.org/10.1038/s41524-021-00498-5} {\bibfield  {journal} {\bibinfo  {journal} {npj Comput. Mater.}\ }\textbf {\bibinfo {volume} {7}},\ \bibinfo {pages} {33} (\bibinfo {year} {2021})}\BibitemShut {NoStop}%
\bibitem [{\citenamefont {Scheidemantel}\ \emph {et~al.}(2003)\citenamefont {Scheidemantel}, \citenamefont {Ambrosch-Draxl}, \citenamefont {Thonhauser}, \citenamefont {Badding},\ and\ \citenamefont {Sofo}}]{drude2}%
  \BibitemOpen
  \bibfield  {author} {\bibinfo {author} {\bibfnamefont {T.~J.}\ \bibnamefont {Scheidemantel}}, \bibinfo {author} {\bibfnamefont {C.}~\bibnamefont {Ambrosch-Draxl}}, \bibinfo {author} {\bibfnamefont {T.}~\bibnamefont {Thonhauser}}, \bibinfo {author} {\bibfnamefont {J.~V.}\ \bibnamefont {Badding}},\ and\ \bibinfo {author} {\bibfnamefont {J.~O.}\ \bibnamefont {Sofo}},\ }\bibfield  {title} {\bibinfo {title} {Transport coefficients from first-principles calculations},\ }\href {https://doi.org/10.1103/PhysRevB.68.125210} {\bibfield  {journal} {\bibinfo  {journal} {Phys. Rev. B}\ }\textbf {\bibinfo {volume} {68}},\ \bibinfo {pages} {125210} (\bibinfo {year} {2003})}\BibitemShut {NoStop}%
\bibitem [{\citenamefont {\ifmmode~\check{Z}\else \v{Z}\fi{}elezn\'y}\ \emph {et~al.}(2021)\citenamefont {\ifmmode~\check{Z}\else \v{Z}\fi{}elezn\'y}, \citenamefont {Fang}, \citenamefont {Olejn\'{\i}k}, \citenamefont {Patchett}, \citenamefont {Gerhard}, \citenamefont {Gould}, \citenamefont {Molenkamp}, \citenamefont {Gomez-Olivella}, \citenamefont {Zemen}, \citenamefont {Tich\'y}, \citenamefont {Jungwirth},\ and\ \citenamefont {Ciccarelli}}]{drude3}%
  \BibitemOpen
  \bibfield  {author} {\bibinfo {author} {\bibfnamefont {J.}~\bibnamefont {\ifmmode~\check{Z}\else \v{Z}\fi{}elezn\'y}}, \bibinfo {author} {\bibfnamefont {Z.}~\bibnamefont {Fang}}, \bibinfo {author} {\bibfnamefont {K.}~\bibnamefont {Olejn\'{\i}k}}, \bibinfo {author} {\bibfnamefont {J.}~\bibnamefont {Patchett}}, \bibinfo {author} {\bibfnamefont {F.}~\bibnamefont {Gerhard}}, \bibinfo {author} {\bibfnamefont {C.}~\bibnamefont {Gould}}, \bibinfo {author} {\bibfnamefont {L.~W.}\ \bibnamefont {Molenkamp}}, \bibinfo {author} {\bibfnamefont {C.}~\bibnamefont {Gomez-Olivella}}, \bibinfo {author} {\bibfnamefont {J.}~\bibnamefont {Zemen}}, \bibinfo {author} {\bibfnamefont {T.}~\bibnamefont {Tich\'y}}, \bibinfo {author} {\bibfnamefont {T.}~\bibnamefont {Jungwirth}},\ and\ \bibinfo {author} {\bibfnamefont {C.}~\bibnamefont {Ciccarelli}},\ }\bibfield  {title} {\bibinfo {title} {Unidirectional magnetoresistance and spin-orbit torque in nimnsb},\ }\href {https://doi.org/10.1103/PhysRevB.104.054429} {\bibfield  {journal}
  {\bibinfo  {journal} {Phys. Rev. B}\ }\textbf {\bibinfo {volume} {104}},\ \bibinfo {pages} {054429} (\bibinfo {year} {2021})}\BibitemShut {NoStop}%
\bibitem [{\citenamefont {Zhao}\ \emph {et~al.}(2024)\citenamefont {Zhao}, \citenamefont {Tao}, \citenamefont {Fu}, \citenamefont {Bellaiche},\ and\ \citenamefont {Ma}}]{drude4}%
  \BibitemOpen
  \bibfield  {author} {\bibinfo {author} {\bibfnamefont {H.~J.}\ \bibnamefont {Zhao}}, \bibinfo {author} {\bibfnamefont {L.}~\bibnamefont {Tao}}, \bibinfo {author} {\bibfnamefont {Y.}~\bibnamefont {Fu}}, \bibinfo {author} {\bibfnamefont {L.}~\bibnamefont {Bellaiche}},\ and\ \bibinfo {author} {\bibfnamefont {Y.}~\bibnamefont {Ma}},\ }\bibfield  {title} {\bibinfo {title} {General theory for longitudinal nonreciprocal charge transport},\ }\href {https://doi.org/10.1103/PhysRevLett.133.096802} {\bibfield  {journal} {\bibinfo  {journal} {Phys. Rev. Lett.}\ }\textbf {\bibinfo {volume} {133}},\ \bibinfo {pages} {096802} (\bibinfo {year} {2024})}\BibitemShut {NoStop}%
\bibitem [{Note3()}]{Note3}%
  \BibitemOpen
  \bibinfo {note} {We employ the $xyz$ Cartesian coordinate system with $x$, $y$, and $z$ axes being perpendicular to each other. As a convention, we use $\protect \mathbf {x}$, $\protect \mathbf {y}$, and $\protect \mathbf {z}$ to denote the unit vectors along $x$, $y$, and $z$ axes, respectively.}\BibitemShut {Stop}%
\bibitem [{Note4()}]{Note4}%
  \BibitemOpen
  \bibinfo {note} {Our theory is established under the first-order approximation with respect to electric polarization $P$. Beyond such an approximation, this statement is not exactly true. For instance, $\alpha _P = \alpha _0 + \alpha _1 P + \alpha _3 P^3$ and $\beta _P = \alpha _0 + \beta _1 P + \beta _3 P^3$ may yield non-zero $\sigma _{x,x}(P,\mu )-\sigma _{y,y}(P,\mu )$, when $\alpha _1=\beta _1$ and $\alpha _3\protect \neq \beta _3$.}\BibitemShut {Stop}%
\bibitem [{Note5()}]{Note5}%
  \BibitemOpen
  \bibinfo {note} {As shown in Eqs.~(\ref {eq:disppol})--(\ref {eq:condpol3}), neither magnetism nor spin-orbit interaction is essential for polarization-switchable ECA. Magnetism and spin-orbit interaction are thus neglected in our derivations.}\BibitemShut {Stop}%
\bibitem [{poi()}]{point}%
  \BibitemOpen
  \href@noop {} {\bibinfo {title} {Bilbao crystallographic server: Point}},\ \bibinfo {howpublished} {\url{https://www.cryst.ehu.es/rep/point.html}}\BibitemShut {NoStop}%
\bibitem [{\citenamefont {Koster}\ \emph {et~al.}(1963)\citenamefont {Koster}, \citenamefont {Dimmock}, \citenamefont {Wheeler},\ and\ \citenamefont {Statz}}]{koster}%
  \BibitemOpen
  \bibfield  {author} {\bibinfo {author} {\bibfnamefont {G.~F.}\ \bibnamefont {Koster}}, \bibinfo {author} {\bibfnamefont {J.~D.}\ \bibnamefont {Dimmock}}, \bibinfo {author} {\bibfnamefont {R.~G.}\ \bibnamefont {Wheeler}},\ and\ \bibinfo {author} {\bibfnamefont {H.}~\bibnamefont {Statz}},\ }\href@noop {} {\emph {\bibinfo {title} {Properties of the Thirty-Two Point Group}}}\ (\bibinfo  {publisher} {M.I.T. Press},\ \bibinfo {year} {1963})\BibitemShut {NoStop}%
\bibitem [{spa()}]{space}%
  \BibitemOpen
  \href@noop {} {\bibinfo {title} {Bilbao crystallographic server: Generators and general position}},\ \bibinfo {howpublished} {\url{https://www.cryst.ehu.es/cryst/get_gen.html}}\BibitemShut {NoStop}%
\bibitem [{\citenamefont {Hergert}\ and\ \citenamefont {Geilhufe}(2018)}]{gtpack}%
  \BibitemOpen
  \bibfield  {author} {\bibinfo {author} {\bibfnamefont {W.}~\bibnamefont {Hergert}}\ and\ \bibinfo {author} {\bibfnamefont {R.~M.}\ \bibnamefont {Geilhufe}},\ }\href@noop {} {\emph {\bibinfo {title} {{Group Theory in Solid State Physics and Photonics: Problem Solving with Mathematica}}}}\ (\bibinfo  {publisher} {Wiley-VCH},\ \bibinfo {year} {2018})\BibitemShut {NoStop}%
\bibitem [{\citenamefont {El-Batanouny}\ and\ \citenamefont {Wooten}(2008)}]{symmetry}%
  \BibitemOpen
  \bibfield  {author} {\bibinfo {author} {\bibfnamefont {M.}~\bibnamefont {El-Batanouny}}\ and\ \bibinfo {author} {\bibfnamefont {F.}~\bibnamefont {Wooten}},\ }\href@noop {} {\emph {\bibinfo {title} {Symmetry and Condensed Matter Physics: A Computational Approach}}}\ (\bibinfo  {publisher} {Cambridge University Press},\ \bibinfo {year} {2008})\BibitemShut {NoStop}%
\bibitem [{Note6()}]{Note6}%
  \BibitemOpen
  \bibinfo {note} {By higher-order corrections, we mean corrections that go beyond effective mass approximation or involve higher-order contributions from electric polarization. Such higher-order corrections should be compatible with the corresponding point group symmetry.}\BibitemShut {Stop}%
\bibitem [{Note7()}]{Note7}%
  \BibitemOpen
  \bibinfo {note} {Tables~\ref {tab:pointeca} and~\ref {tab:ferroeca} also contain cases associated with approximately polarization-switchable ECA. We demonstrate and verify this by computing the electrical conductivities for the NaSrP semiconductor, which is shown in~\protect \textit {Section D} of our SM.}\BibitemShut {Stop}%
\bibitem [{\citenamefont {Richman}(1968)}]{alpcryst}%
  \BibitemOpen
  \bibfield  {author} {\bibinfo {author} {\bibfnamefont {D.}~\bibnamefont {Richman}},\ }\bibfield  {title} {\bibinfo {title} {Vapor phase growth and properties of aluminum phosphide},\ }\href {https://doi.org/10.1149/1.2411483} {\bibfield  {journal} {\bibinfo  {journal} {J. Electrochem. Soc.}\ }\textbf {\bibinfo {volume} {115}},\ \bibinfo {pages} {945} (\bibinfo {year} {1968})}\BibitemShut {NoStop}%
\bibitem [{\citenamefont {Hinuma}\ \emph {et~al.}(2014)\citenamefont {Hinuma}, \citenamefont {Gr\"uneis}, \citenamefont {Kresse},\ and\ \citenamefont {Oba}}]{zinckresse}%
  \BibitemOpen
  \bibfield  {author} {\bibinfo {author} {\bibfnamefont {Y.}~\bibnamefont {Hinuma}}, \bibinfo {author} {\bibfnamefont {A.}~\bibnamefont {Gr\"uneis}}, \bibinfo {author} {\bibfnamefont {G.}~\bibnamefont {Kresse}},\ and\ \bibinfo {author} {\bibfnamefont {F.}~\bibnamefont {Oba}},\ }\bibfield  {title} {\bibinfo {title} {Band alignment of semiconductors from density-functional theory and many-body perturbation theory},\ }\href {https://doi.org/10.1103/PhysRevB.90.155405} {\bibfield  {journal} {\bibinfo  {journal} {Phys. Rev. B}\ }\textbf {\bibinfo {volume} {90}},\ \bibinfo {pages} {155405} (\bibinfo {year} {2014})}\BibitemShut {NoStop}%
\bibitem [{\citenamefont {Gr\"uneis}\ \emph {et~al.}(2014)\citenamefont {Gr\"uneis}, \citenamefont {Kresse}, \citenamefont {Hinuma},\ and\ \citenamefont {Oba}}]{zinckresse2}%
  \BibitemOpen
  \bibfield  {author} {\bibinfo {author} {\bibfnamefont {A.}~\bibnamefont {Gr\"uneis}}, \bibinfo {author} {\bibfnamefont {G.}~\bibnamefont {Kresse}}, \bibinfo {author} {\bibfnamefont {Y.}~\bibnamefont {Hinuma}},\ and\ \bibinfo {author} {\bibfnamefont {F.}~\bibnamefont {Oba}},\ }\bibfield  {title} {\bibinfo {title} {Ionization potentials of solids: The importance of vertex corrections},\ }\href {https://doi.org/10.1103/PhysRevLett.112.096401} {\bibfield  {journal} {\bibinfo  {journal} {Phys. Rev. Lett.}\ }\textbf {\bibinfo {volume} {112}},\ \bibinfo {pages} {096401} (\bibinfo {year} {2014})}\BibitemShut {NoStop}%
\bibitem [{\citenamefont {Ioanid}\ \emph {et~al.}(1984)\citenamefont {Ioanid}, \citenamefont {Popescu}, \citenamefont {Vlahovici},\ and\ \citenamefont {Bunget}}]{kdp1}%
  \BibitemOpen
  \bibfield  {author} {\bibinfo {author} {\bibfnamefont {A.}~\bibnamefont {Ioanid}}, \bibinfo {author} {\bibfnamefont {M.}~\bibnamefont {Popescu}}, \bibinfo {author} {\bibfnamefont {N.}~\bibnamefont {Vlahovici}},\ and\ \bibinfo {author} {\bibfnamefont {I.}~\bibnamefont {Bunget}},\ }\bibfield  {title} {\bibinfo {title} {On the phase transition of {KH\textsubscript{2}PO\textsubscript{4}} at {$T \approx 110\,^\circ\mathrm{C}$}},\ }\href {https://doi.org/10.1002/pssa.2210820235} {\bibfield  {journal} {\bibinfo  {journal} {Phys. Status Solidi A}\ }\textbf {\bibinfo {volume} {82}},\ \bibinfo {pages} {K125} (\bibinfo {year} {1984})}\BibitemShut {NoStop}%
\bibitem [{\citenamefont {Jia}\ \emph {et~al.}(2020)\citenamefont {Jia}, \citenamefont {Cheng}, \citenamefont {Whangbo}, \citenamefont {Hong},\ and\ \citenamefont {Deng}}]{kdp2}%
  \BibitemOpen
  \bibfield  {author} {\bibinfo {author} {\bibfnamefont {M.}~\bibnamefont {Jia}}, \bibinfo {author} {\bibfnamefont {X.}~\bibnamefont {Cheng}}, \bibinfo {author} {\bibfnamefont {M.-H.}\ \bibnamefont {Whangbo}}, \bibinfo {author} {\bibfnamefont {M.}~\bibnamefont {Hong}},\ and\ \bibinfo {author} {\bibfnamefont {S.}~\bibnamefont {Deng}},\ }\bibfield  {title} {\bibinfo {title} {Second harmonic generation responses of {KH\textsubscript{2}PO\textsubscript{4}}: importance of k and breaking down of kleinman symmetry},\ }\href {https://doi.org/10.1039/d0ra03136d} {\bibfield  {journal} {\bibinfo  {journal} {RSC Adv.}\ }\textbf {\bibinfo {volume} {10}},\ \bibinfo {pages} {26479} (\bibinfo {year} {2020})}\BibitemShut {NoStop}%
\bibitem [{\citenamefont {Miyoshi}\ \emph {et~al.}(2011)\citenamefont {Miyoshi}, \citenamefont {Mashiyama}, \citenamefont {Asahi}, \citenamefont {Kimura},\ and\ \citenamefont {Noda}}]{kdp3}%
  \BibitemOpen
  \bibfield  {author} {\bibinfo {author} {\bibfnamefont {T.}~\bibnamefont {Miyoshi}}, \bibinfo {author} {\bibfnamefont {H.}~\bibnamefont {Mashiyama}}, \bibinfo {author} {\bibfnamefont {T.}~\bibnamefont {Asahi}}, \bibinfo {author} {\bibfnamefont {H.}~\bibnamefont {Kimura}},\ and\ \bibinfo {author} {\bibfnamefont {Y.}~\bibnamefont {Noda}},\ }\bibfield  {title} {\bibinfo {title} {Single-crystal neutron structural analyses of potassium dihydrogen phosphate and potassium dideuterium phosphate},\ }\href {https://doi.org/10.1143/jpsj.80.044709} {\bibfield  {journal} {\bibinfo  {journal} {J. Phys. Soc. Jpn.}\ }\textbf {\bibinfo {volume} {80}},\ \bibinfo {pages} {044709} (\bibinfo {year} {2011})}\BibitemShut {NoStop}%
\bibitem [{\citenamefont {Hestroffer}\ \emph {et~al.}(2018)\citenamefont {Hestroffer}, \citenamefont {Sperlich}, \citenamefont {Dadgostar}, \citenamefont {Golz}, \citenamefont {Krumland}, \citenamefont {Masselink},\ and\ \citenamefont {Hatami}}]{alpdopingbe}%
  \BibitemOpen
  \bibfield  {author} {\bibinfo {author} {\bibfnamefont {K.}~\bibnamefont {Hestroffer}}, \bibinfo {author} {\bibfnamefont {D.}~\bibnamefont {Sperlich}}, \bibinfo {author} {\bibfnamefont {S.}~\bibnamefont {Dadgostar}}, \bibinfo {author} {\bibfnamefont {C.}~\bibnamefont {Golz}}, \bibinfo {author} {\bibfnamefont {J.}~\bibnamefont {Krumland}}, \bibinfo {author} {\bibfnamefont {W.~T.}\ \bibnamefont {Masselink}},\ and\ \bibinfo {author} {\bibfnamefont {F.}~\bibnamefont {Hatami}},\ }\bibfield  {title} {\bibinfo {title} {Transport properties of doped {A}l{P} for the development of conductive {A}l{P}/{G}a{P} distributed bragg reflectors and their integration into light-emitting diodes},\ }\href {http://dx.doi.org/10.1063/1.5024632} {\bibfield  {journal} {\bibinfo  {journal} {Appl. Phys. Lett.}\ }\textbf {\bibinfo {volume} {112}},\ \bibinfo {pages} {192107} (\bibinfo {year} {2018})}\BibitemShut {NoStop}%
\bibitem [{\citenamefont {Zhu}\ and\ \citenamefont {Park}(2006)}]{mtjmr2}%
  \BibitemOpen
  \bibfield  {author} {\bibinfo {author} {\bibfnamefont {J.-G.~J.}\ \bibnamefont {Zhu}}\ and\ \bibinfo {author} {\bibfnamefont {C.}~\bibnamefont {Park}},\ }\bibfield  {title} {\bibinfo {title} {Magnetic tunnel junctions},\ }\href {https://doi.org/10.1016/s1369-7021(06)71693-5} {\bibfield  {journal} {\bibinfo  {journal} {Mater. Today}\ }\textbf {\bibinfo {volume} {9}},\ \bibinfo {pages} {36} (\bibinfo {year} {2006})}\BibitemShut {NoStop}%
\bibitem [{\citenamefont {Qin}\ \emph {et~al.}(2023)\citenamefont {Qin}, \citenamefont {Yan}, \citenamefont {Wang}, \citenamefont {Chen}, \citenamefont {Meng}, \citenamefont {Dong}, \citenamefont {Zhu}, \citenamefont {Cai}, \citenamefont {Feng}, \citenamefont {Zhou}, \citenamefont {Liu}, \citenamefont {Zhang}, \citenamefont {Zeng}, \citenamefont {Zhang}, \citenamefont {Jiang},\ and\ \citenamefont {Liu}}]{mtjmr}%
  \BibitemOpen
  \bibfield  {author} {\bibinfo {author} {\bibfnamefont {P.}~\bibnamefont {Qin}}, \bibinfo {author} {\bibfnamefont {H.}~\bibnamefont {Yan}}, \bibinfo {author} {\bibfnamefont {X.}~\bibnamefont {Wang}}, \bibinfo {author} {\bibfnamefont {H.}~\bibnamefont {Chen}}, \bibinfo {author} {\bibfnamefont {Z.}~\bibnamefont {Meng}}, \bibinfo {author} {\bibfnamefont {J.}~\bibnamefont {Dong}}, \bibinfo {author} {\bibfnamefont {M.}~\bibnamefont {Zhu}}, \bibinfo {author} {\bibfnamefont {J.}~\bibnamefont {Cai}}, \bibinfo {author} {\bibfnamefont {Z.}~\bibnamefont {Feng}}, \bibinfo {author} {\bibfnamefont {X.}~\bibnamefont {Zhou}}, \bibinfo {author} {\bibfnamefont {L.}~\bibnamefont {Liu}}, \bibinfo {author} {\bibfnamefont {T.}~\bibnamefont {Zhang}}, \bibinfo {author} {\bibfnamefont {Z.}~\bibnamefont {Zeng}}, \bibinfo {author} {\bibfnamefont {J.}~\bibnamefont {Zhang}}, \bibinfo {author} {\bibfnamefont {C.}~\bibnamefont {Jiang}},\ and\ \bibinfo {author} {\bibfnamefont {Z.}~\bibnamefont {Liu}},\ }\bibfield  {title} {\bibinfo {title}
  {Room-temperature magnetoresistance in an all-antiferromagnetic tunnel junction},\ }\href {https://doi.org/10.1038/s41586-022-05461-y} {\bibfield  {journal} {\bibinfo  {journal} {Nature}\ }\textbf {\bibinfo {volume} {613}},\ \bibinfo {pages} {485} (\bibinfo {year} {2023})}\BibitemShut {NoStop}%
\end{thebibliography}

\end{document}